\documentclass[twocolumn]{aastex631}

\usepackage{float}
\usepackage{verbatim}
\usepackage{soul}
\usepackage{hyperref}

\usepackage{booktabs,chemformula}
\shorttitle{DWD Binaries in SDSS-V DR19}
\shortauthors{Adamane Pallathadka et al.}

\graphicspath{{./}{fig/}}

\begin{document}

\title{Double White Dwarf Binaries in SDSS-V DR19 : A catalog of DA white dwarf binaries and constraints on the binary population}

\author[0000-0002-5864-1332]{Gautham Adamane Pallathadka}
\affiliation{William H. Miller III Department of
Physics \& Astronomy, Johns Hopkins University, 3400 N Charles St, Baltimore, MD 21218, USA}

\author[0000-0002-0572-8012]{Vedant Chandra}
\affiliation{Center for Astrophysics $\mid$ Harvard \& Smithsonian, 60 Garden St., Cambridge, MA 02138, USA}

\author[0000-0001-6100-6869]{Nadia L. Zakamska}
\affiliation{William H. Miller III Department of
Physics \& Astronomy, Johns Hopkins University, 3400 N Charles St, Baltimore, MD 21218, USA}

\author[0000-0002-8866-4797]{Nicole R. Crumpler}
\affiliation{William H. Miller III Department of
Physics \& Astronomy, Johns Hopkins University, 3400 N Charles St, Baltimore, MD 21218, USA}

\author[0000-0002-6270-8624]{Stefan M. Arseneau}
\affiliation{William H. Miller III Department of
Physics \& Astronomy, Johns Hopkins University, 3400 N Charles St, Baltimore, MD 21218, USA}
\affiliation{Department of Astronomy \& Institute for Astrophysical Research, Boston University, 725 Commonwealth Ave., Boston, MA 02215, USA}

\author[0000-0002-6871-1752]{Kareem El-Badry}
\affiliation{Department of Astronomy, California Institute of Technology, 1200 East California Boulevard, Pasadena, CA 91125, USA}

\author[0000-0002-2761-3005]{Boris T. G\"{a}nsicke}
\affiliation{Department of Physics, University of Warwick, Coventry CV4 7AL, UK}

\author[0000-0002-0632-8897]{Yossef Zenati}
\affiliation{William H. Miller III Department of
Physics \& Astronomy, Johns Hopkins University, 3400 N Charles St, Baltimore, MD 21218, USA}
\affiliation{Space Telescope Science Institute, Baltimore, MD 21218, USA}
\affiliation{Astrophysics Research Center of the Open University (ARCO), The Open University of Israel, Ra’anana 4353701, Israel}
\affiliation{Department of Natural Sciences, The Open University of Israel, Ra’anana 4353701, Israel}

\author[0000-0001-5941-2286]{J.J. Hermes}
\affiliation{Department of Astronomy \& Institute for Astrophysical Research, Boston University, 725 Commonwealth Ave., Boston, MA 02215, USA}

\author[0000-0003-3441-9355]{Axel D. Schwope}
\affiliation{Leibniz-Institut für Astrophysik Potsdam (AIP), An der Sternwarte 16, 14482 Potsdam, Germany }

\author[0000-0003-3494-343X]{Carles Badenes}
\affiliation{Department of Physics and Astronomy, University of Pittsburgh, 3941 O'Hara Street, Pittsburgh, PA 15260, USA}
\affiliation{Pittsburgh Particle Physics, Astrophysics, and Cosmology Center (PITT PACC), University of Pittsburgh, Pittsburgh, PA 15260, USA}

\author[0000-0002-6428-4378]{Nicola Pietro Gentile Fusillo}
\affiliation{Department of Physics, Universita’ degli Studi di Trieste, Via A. Valerio 2, 34127, Trieste, Italy}

\author[0000-0002-6770-2627]{Sean Morrison}
\affiliation{Department of Astronomy, University of Illinois at Urbana-Champaign, Urbana, IL 61801, USA}

\author[0000-0001-7296-3533]{Tim Cunningham}
\affiliation{Center for Astrophysics $\mid$ Harvard \& Smithsonian, 60 Garden St., Cambridge, MA 02138, USA}

\author[0000-0002-4469-2518]{Priyanka Chakraborty}
\affiliation{Center for Astrophysics $\mid$ Harvard \& Smithsonian, 60 Garden St., Cambridge, MA 02138, USA}

\author[0000-0002-2953-7528]{Gagik Tovmasian}
\affiliation{Instituto de Astronomıa, Universidad Nacional Autónoma de México, A.P. 70-264, 04510, Mexico, D.F., México}

\author[0000-0002-3601-133X]{Dmitry Bizyaev}
\affiliation{Apache Point Observatory, P.O. Box 59, Sunspot, NM 88349}

\author[0000-0002-2835-2556]{Kaike Pan}
\affiliation{Apache Point Observatory, P.O. Box 59, Sunspot, NM 88349}

\author[0000-0002-6404-9562]{Scott F. Anderson}
\affiliation{Department of Astronomy, University of Washington, Box 351580, Seattle, WA 98195, USA}

\author[0009-0006-8478-7163]{Sebastian Demasi }
\affiliation{Department of Astronomy, University of Washington, Box 351580, Seattle, WA 98195, USA}

\correspondingauthor{Gautham Adamane Pallathadka}
\email{gadaman1@jh.edu}

\begin{abstract}

\noindent
The fifth--generation Sloan Digital Sky Survey (SDSS-V) includes the first large-scale spectroscopic survey of white dwarfs (WDs) in the era of Gaia parallaxes. SDSS-V collects multiple exposures per target, making it ideal for binary detection.
We present a search for hydrogen atmosphere (DA) double white dwarf (DWD) binaries in this rich dataset. We quantify radial velocity variations between sub-exposures to identify binary candidates, and also measure the orbital period for a subset of DWD binary candidates. We find 63 DWD binary candidates, of which 43 are new discoveries, and we provide tentative periods for 10 binary systems. Using these measurements, we place constraints on the binary fraction of the Galactic WD population with $< 0.4$~AU separations $f_{\mathrm{bin,0.4}} = 9\%$, and the power-law index of the initial  separation distribution $\alpha = -0.62$. 
Using the simulated binary population, we estimate that $\leq 10$ super-Chandrasekhar binaries that merge within a Hubble time are expected in our sample. We predict that $\leq 5$ systems in our sample should be detectable via gravitational waves by LISA (Laser Interferometer Space Antenna), one of which has already been identified as a LISA verification source. We also estimate a total of about 10\,000 -- 20\,000 LISA-detectable DWD binaries in the galaxy. Our catalog of WD+WD binary candidates in SDSS-V is now public, and promises to uncover a large number of exciting DWD systems. 
\end{abstract}


\section{Introduction} \label{sec:intro}

Double white dwarf (DWD) systems are compact binaries of two white dwarfs (WD+WD) that may be a major source of Type Ia supernovae (SNeIa; see \citealt{maoz_observational_2014} for a review), cosmological standard candles used to measure the accelerating expansion of the Universe \citep{riess_observational_1998,perlmutter_measurements_1999}. The DWDs with the shortest period will be dominant gravitational wave sources in the millihertz range detectable by space-based observatories \citep{marsh_double_2011,kupfer_lisa_2018,lamberts_predicting_2019,li_gravitational-wave_2020}. Studying the population of double degenerates across a wide range of periods and masses contributes to our understanding of binary evolution from common envelopes to mergers \citep{nelemans_reconstructing_2000,maxted_mass_2002,marsh_mass_2004,van_der_sluys_modelling_2006,brown_most_2016,inight_towards_2021}. 

A DWD binary with a period less than $\approx$ 10 hours can merge due to gravitational wave emission within Hubble time ($\approx$ 13 Gyrs), and if the total mass is greater than the Chandrasekhar limit ($>$1.4 M$_{\odot}$) it can lead to Type-Ia supernova. Recently, it has been shown that even sub-Chandrasekhar mass WD explosion may also lead to Type-Ia supernova for certain binary configurations \citep{shen_every_2015,shen_sub-chandrasekhar-mass_2018,shen_almost_2024}. The quest to understand the origin of Type Ia supernovae has motivated several searches for DWD binaries, but despite decades of research, there has been no unambiguous discovery of a progenitor with a total mass greater than $>$1.4 M$_{\odot}$ \citep{maoz_type-ia_2012}.
Until the early 2000s, about 10-15 DWD binaries were known with full orbital solutions \citep{maxted_radial_2002,nelemans_population_2001,maxted_mass_2002}. These sources were used to calibrate binary population synthesis models, to elucidate close binary evolution processes, and to estimate the viability of the double degenerate model of Type-Ia supernovae. Since then, through dedicated searches such as ESO SPY (SN Ia Progenitor surveY) survey \citep{napiwotzki_binaries_2002,napiwotzki_eso_2020}, SWARMS survey \citep{badenes_first_2009}, ELM survey \citep{brown_elm_2010,brown_elm_2020}, DBL survey \citep{munday_dbl_2024}, and through serendipitous discoveries \citep{burdge_88_2020,kilic_hidden_2021}, the number of known DWD binaries have increased to approximately 300 systems.

In contrast to the mass distribution of single WDs which peaks at 0.6 M$_{\odot}$ \citep{kepler_white_2019,crumpler_large_2025}, the mass distribution of known DWD binaries leans towards lower massed WDs. This is primarily because a large fraction of known DWDs are extremely low mass (ELM) WDs with mass less than 0.4 M$_{\odot}$, which were specifically targeted by the ELM survey as they are particularly interesting because of their origin in close stellar binaries. Additionally, binaries of low-mass WDs are brighter and are thus more likely to be serendipitously discovered. 

There have been no unambiguous Type-Ia progenitors discovered so far with merger time less than the age of the universe. Two very promising systems SDSS J075141.18-014120.9 and  SDSS J174140.49+652638.7, have orbital periods of 1.9 hrs and 1.5 hrs, and total mass of 1.16 M$_{\odot}$ and $>$1.28 M$_{\odot}$, respectively \citep{kilic_found_2013}. However, the companion masses and core compositions remain uncertain, and owing to large mass ratios these binaries may lead to faint Type .Ia supernovae instead. SDSS J213228.36+075428.2 is a similar system which is not expected to lead to Type-Ia supernova despite a large total mass \citep{kilic_massive_2016}. 
SBS 1150+599A is another system which could potentially be a Type-Ia supernova progenitor \citep{tovmassian_double-degenerate_2010}. The total mass of the system may be greater than the Chandrasekhar limit. However, the low mass star in this system is currently bloated, due to the recent nebular phase, and the large uncertainty in its mass makes it difficult to confirm its nature as a Type-Ia progenitor. J1138-5139 is another recently discovered Type-Ia progenitor candidate with a total system mass 1.27 M$_{\odot}$ \citep{kosakowski_new_2025}. However, as discussed by \cite{chickles_gravitational-wave-detectable_2025}, the future of this system is not yet clear and it may also result in a Type .Ia supernova or evolve into an AM CVn system. Recently, \cite{munday_super-chandrasekhar_2025} reported the discovery of a DWD binary with a mass greater than the Chandrasekhar limit with a merger time of 22 Gyrs, tantalizingly close to Hubble time. This system is currently the closest we have come to the discovery of a Type-Ia progenitor with a mass greater than 1.4 M$_{\odot}$ that can merge within Hubble time.

Despite the dramatic increase in the number of known DWDs in recent years, the total number of DWDs is still small. The total number of spectroscopically confirmed WDs is over 30\,000, while the number of confirmed DWDs is only about 300 \citep{munday_dbl_2024}\footnote{https://github.com/JamesMunday98/CloseDWDbinaries}. This is about 1\% despite the estimated DWD binary fraction being about 10\% for binaries with separations within 4~AU \citep{badenes_merger_2012,maoz_binary_2017,maoz_separation_2018}. This is primarily because WDs are faint, and have broad absorption lines due to pressure broadening. Furthermore, the vast majority of WDs only have a couple of high-quality exposures. Thus, measuring radial velocity (RV) to high precision is difficult and the lack of the large number of exposures makes the binary search a challenging task.

The new generation of Sloan Digital Sky Survey \citep[SDSS-V;][]{kollmeier_sloan_2025} is an all-sky survey that aims to explicitly observe up to $60\,000$ white dwarfs. Each co-added SDSS-V spectrum is composed of numerous 15-minute sub-exposures taken consecutively or split over multiple nights. This makes it a promising setup to find new DWD binaries by monitoring and quantifying the RV variation in the visible brighter star as it moves in its orbit \citep{badenes_first_2009,breedt_using_2017,chandra_99-minute_2021,adamane_pallathadka_discovery_2024,adamane_pallathadka_double_2025}. 

The DWD binary candidates selected by RV monitoring are, typically, unequal mass binaries with the visible WD being the least massive of the two. WDs are supported by electron degeneracy pressure, and therefore have a mass-radius relationship such that less massive WDs have larger radii and are therefore typically brighter \citep{chandrasekhar_maximum_1931,arseneau_measuring_2024,crumpler_detection_2024}. In binaries, this causes the massive companion to remain invisible while the less massive WD shows RV variations. In equal mass binaries, usually both WDs have comparable size and brightness, and the absorption lines become blended. These `double-lined' binaries are more challenging to identify with low-resolution spectroscopy \citep{chandra_99-minute_2021}. \cite{munday_dbl_2024} present a catalog of such double lined systems detected through their luminosity in the color-magnitude diagram.

In this paper, we present a catalog of 63 DWD binary candidates identified by measuring the RV variation across different SDSS-V sub-exposures. Using this binary catalog, we estimate the properties of underlying DWD binary population and estimate the expected number of Type-Ia progenitors in the sample and the number of LISA detectable binaries. In Sec.~\ref{sec:Data}, we summarise the SDSS-V dataset and present our selection cuts to build the WD sample analyzed in this paper. In Sec.~\ref{sec:Analysis}, we present our analysis used to identify binary candidates, and in Sec.~\ref{sec:catalog} we present our catalog of DWD binary candidates. In Sec.~\ref{sec:binary_population}, we derive constraints on the DWD binary population. Finally, in Sec.~\ref{sec:Discussion} we summarize our findings.

\section{Data \label{sec:Data}}



The fifth generation of the Sloan Digital Sky Survey is an ongoing all-sky survey that aims to spectroscopically observe up to 60\,000 WDs. The survey began in November 2020, initially using the 2.5 m telescope at the Apache Point Observatory \citep{gunn_25_2006} and has now expanded to the 2.5 m telescope at Las Campanas Observatory  \citep{bowen_optical_1973}, and observes targets using the Baryon Oscillation Spectroscopic Survey spectrograph (BOSS; \citealt{smee_multi-object_2013}) and the Apache Point Observatory Galactic Evolution Experiment (APOGEE) spectrographs \citep{wilson_apache_2019}. Most WDs are observed as a part of the multi-epoch Milky Way mapper program (Kollmeier et al. 2025, submitted). Each SDSS-V sub-exposure takes 900s and has a median resolution of R$\sim$ 1800, and covers the wavelength range from 3600 \AA\ to 10000 \AA.

As of SDSS-V DR 19 \citep{sdss_collaboration_nineteenth_2025}, over 120\,000 WD sub-exposures have been collected for 19\,000 WDs and , and this is our parent sample. All the data were reduced using internal pipeline \verb|IDLspec2D v6_1_3|. SDSS spectra on the vacum wavelength scale are corrected to the heliocentric frame, and the wavelength calibration is accurate to within $<$10 km s$^{-1}$ \citep{crumpler_large_2025}. 
The previous generations of SDSS observed WDs serendipitously by targeting them initially as quasars, which results in a biased sample \citep{crumpler_large_2025}. SDSS-V aims for a complete sample of WD population by selecting the targets from Gaia sample of WDs by \cite{gentilefusillo_gaia_2019,gentile_fusillo_catalogue_2021}. In SDSS-V, the observations are divided into different observation cartons that select the targets based on science goals e.g WD observations under \texttt{mwm\_wd\_core}, compact binary candidate observations under \texttt{mwm\_cb}, with some targets assigned to multiple cartons \citep{sdss_collaboration_nineteenth_2025,kollmeier_sloan_2025}. Each carton has a different priority, which increases the likelihood of observation of targets assigned under that carton. All cartons which have WD observations have similar priorities and thus equally likely to be observed \citep{almeida_eighteenth_2023}. We also find that the number of exposures for WDs assigned under binary candidate cartons and nonbinary candidate cartons have nearly identical distribution. Thus, SDSS-V gives us an unbiased sample of WDs that is ideal for population studies.

Each WD exposure is classified using a machine learning algorithm, included in the SDSS-V pipeline, named SnowWhite. SnowWhite uses a random forest classifier that is trained on up to 30\,000 WD spectra from SDSS data releases until DR14. These WD spectra were originally classified into 26 possible WD subclasses by visual inspection. Cool WDs with T$_{\mathrm{eff}} <$ 7000 K were underrepresented in older SDSS data releases. To minimize any bias in classification due to this paucity, 3000 model hydrogen atmosphere (DA) WD in the range of 3000 K -- 7000 K and $\log g$ between 7 and 9.5, from \citet{tremblay_spectroscopic_2013}, were added to the training sample. The resolution of model spectra was reduced to match the SDSS resolution and random noise was added to produce SDSS-like spectra with signal-to-noise ratio (SNR) between 5 and 30. 
To ensure a clean sample, we only include DA WDs with spectral signal-to-noise ratio (SNR) $>$ 3 in our catalog. We further remove any WDs that are classified to have greater than 10\% probability to be a featureless DC WD, an extremely hot DO WD, a WD+main-sequence binary, or likely to be a catclysmic Variable (CV) -- all of which lead to spectra that may cause issues in RV measurement. Certain WDs can have different classifications for different exposures. In these cases, only those exposures which satisfy the above criteria are dropped, while the rest are retained. 

\begin{figure*}
    \centering
    \includegraphics[width=\linewidth]{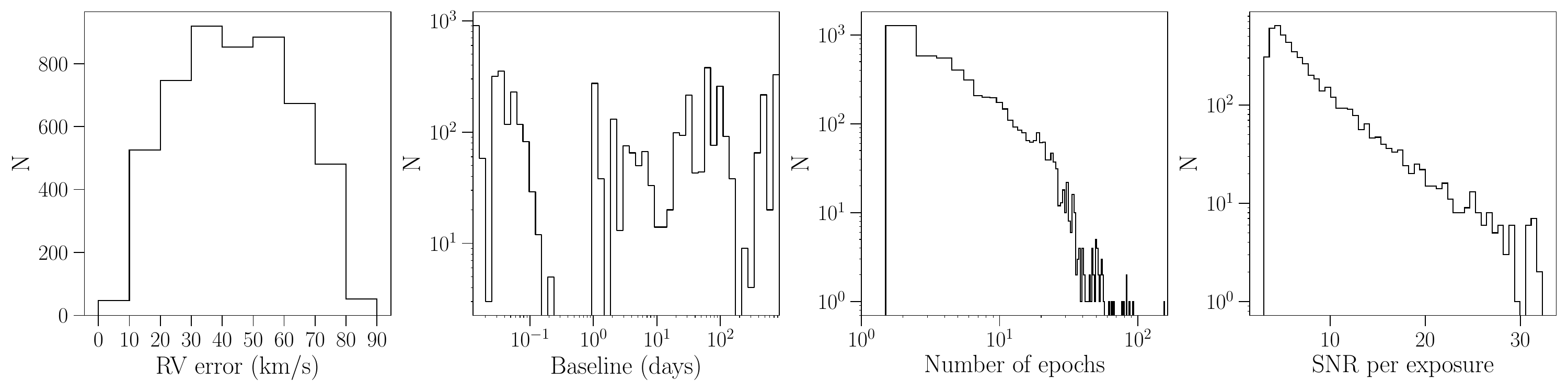}
    \caption{Distribution of median RV error, observational baseline, number of epochs, and median SNR per exposure for each WD is shown.}
    \label{fig:observation_details}
\end{figure*}

We apply further cuts based on the measured RV errors which we describe in Sec.\ref{sec:Analysis}, and keep only those WDs with at least two exposures of data. In Fig.\ref{fig:observation_details}, we summarize the observational details such as distribution of median RV errors, observational baseline per WD, number of epochs per WD, and SNR per exposure of the cleaned sample that is central to our analysis. The median SNR per sub-exposure is $\approx 6$, and the median number of exposures per target in the cleaned sample is 4. The median distance for this sample is 206 pc, and almost 95\% of the sample is within 700 pc. 

\section{Analysis \label{sec:Analysis}}
We analyze the SDSS spectra using CORV \citep[Compact Object Radial Velocity;][]{arseneau_measuring_2024}\footnote{\url{https://github.com/vedantchandra/corv}}. CORV measures the stellar parameters by fitting the atmospheric absorption lines to stellar templates. We fit the first 4 hydrogen Balmer lines -- H$\alpha$, H$\beta$, H$\gamma$, and H$\delta$ to 3D DA WD templates by \cite{tremblay_3d_2015}\footnote{\url{https://warwick.ac.uk/fac/sci/physics/research/astro/people/tremblay/modelgrids/}}. These templates cover a temperature range between 4500 K and 40000 K and $\log g$ between 7 and 9.

The fitting procedure follows two steps -- we first fit the absorption lines using {\tt lmfit} Python package by varying T$_{\mathrm{eff}}$, $\log g$, and the RV, to minimize the $\chi^2$ between the observed spectrum and the template. This gives us a preliminary guess for the parameters. WD spectral templates suffer from well-known degeneracies and can have hot and cold solutions on either side of around 15000K \citep{chandra_computational_2020}. We carry out the initial fit over three different initial temperatures between $5500$ K and $27000$ K to ensure that the parameter space is well explored, and the best-fit is attained. 
The best-fit solution, so obtained, acts as a template for the RV determination which is done by cross-correlating the SDSS spectrum with the template by varying the RV on a fixed finely spaced grid between $-2500$~km s$^{-1}$ and $2500$ km s$^{-1}$. The resulting $\chi^2$ curve is used to obtain the best fit RV and the associated error \citep{arseneau_measuring_2024}. 

Some of the SDSS spectra have wavelength calibration issues. To remove the affected exposures, we fit skylines at $6863.9$ \AA, $7276.4$ \AA, $7340.8$ \AA\ (in air wavelengths) and discard any exposure where all three of them differ from the rest-frame wavelengths by more than $20$ km s$^{-1}$, after applying the heliocentric corrections. Several exposures were polluted by cosmic rays near one of the absorption lines or, rarely, emission lines. This leads to incorrect radial velocity measurement due to incorrect template fitting. To eliminate these, we re-measure the radial velocities of each exposure by fitting H$\beta$, H$\gamma$, and H$\delta$ simultaneously, and then repeat with H$\alpha$ alone. We compare the RVs so obtained and discard any sub exposure where the RVs disagree at greater than 3$\sigma$ level. 

\begin{figure}
    \centering
    \includegraphics[width=1\linewidth]{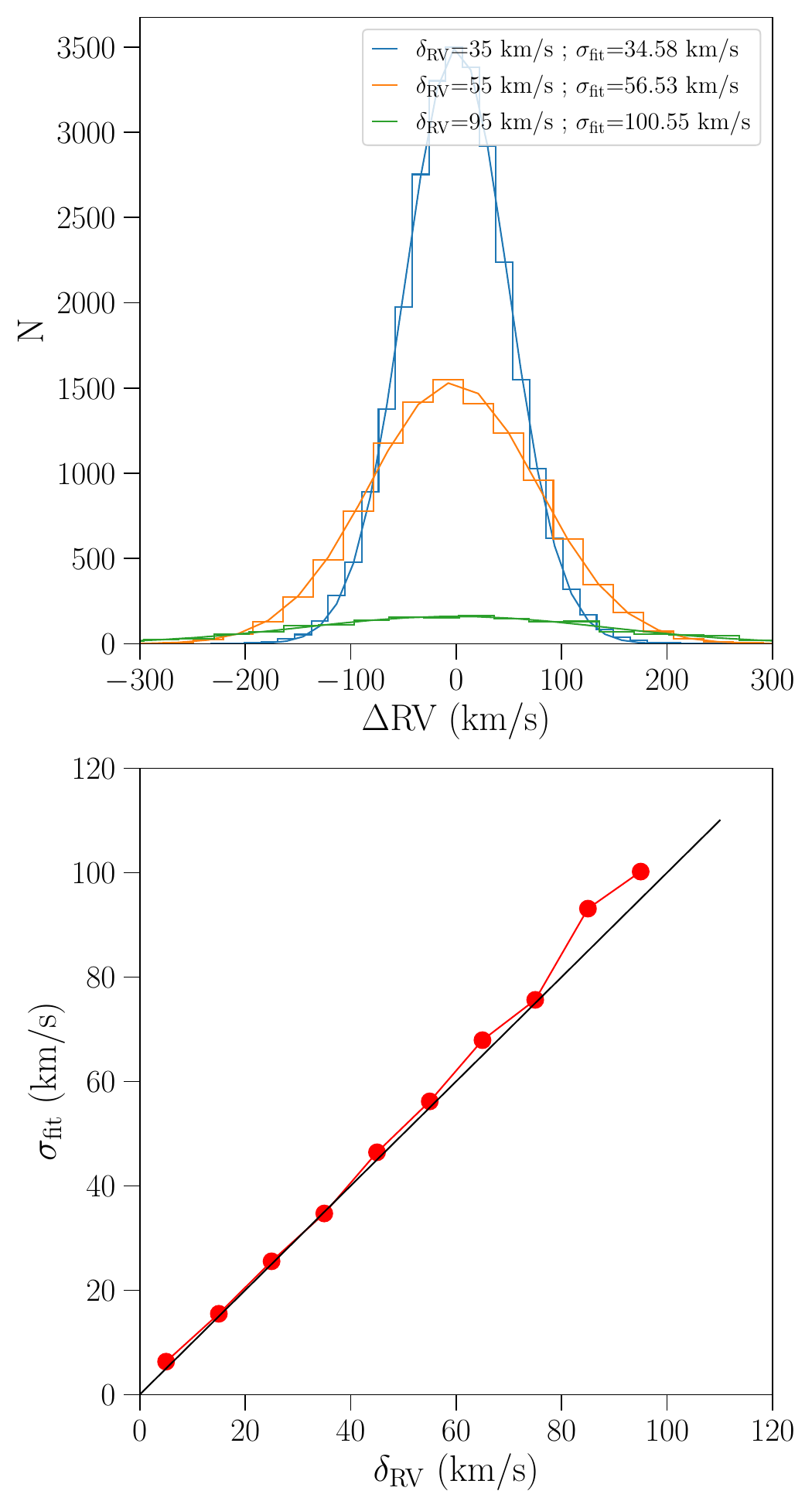}
    \caption{Top: Distribution of pair difference of RVs is shown for three different RV errors. Bottom: The comparison of corrected CORV RV errors and RV errors estimated by pair subtraction is shown. 
    We find that including the correction brings CORV RV errors to within 10\% agreement with the fitted errors from pair-subtraction test.}
    \label{fig:pair_subtract}
\end{figure}

Detection of binaries relies upon accurate measurement of RV errors. To determine the accuracy of RV errors estimated by CORV we perform a pair-subtraction test as outlined by \cite{badenes_merger_2012}. We first group the RV measurements of each WD into ten bins that have similar RV errors. Each bin collects all RV measurements with errors closest to $5$ km s$^{-1}$, $15$ km s$^{-1}$ and so on until $95$ km s$^{-1}$. Within each bin, we calculate the unique pair difference, $\Delta\mathrm{RV}$, between all available RV measurements of each WD. If there is no RV variation, the distribution of the pair differences should follow a Gaussian distribution with a width $\sqrt{2}$ times the RV error of the group, labeled $\delta_{\mathrm{RV}}$. Half the difference between the expected RV error of the group ($5$ km s$^{-1}$, $15$ km s$^{-1}$, etc) and observed RV error of the group from the fit to the distributions is then added to each RV error measurement in that group. This minimizes any deviations in the RV errors measured using CORV. We find that the correction brings measured RV errors to within 10\% agreement with the
fitted errors from pair-subtraction test. This shows that the RVs and RV errors measured using SDSS-V spectra are extremely reliable, which has also been previously noted by \cite{arseneau_measuring_2024}. The results after the correction are shown in Fig.~\ref{fig:pair_subtract}. Finally, we remove the sub-exposures where errors in the RV measurement are greater than $90$ km s$^{-1}$, beyond which the RV errors cannot be reliably calibrated. With these we obtain a sample of $5\,185$ WDs and a total of $42\,176$ sub-exposures.

We make use of the RVs of each sub exposure to look for statistically significant RV variation. We follow the procedures outlined in \citet{maxted_radial_2002} and \citet{breedt_using_2017} -- we calculate $\chi^2_{\mathrm{m}}$, the $\chi^2$ value for the RV variation about the weighted average of all of the RVs for each WD. Under the null hypothesis that the observed RV variation is purely due to spectral noise and not physical motion, $\chi^2_{\mathrm{m}}$ should follow a $\chi^2$ distribution with $(n-1)$ degrees of freedom, where $n$ is the number of sub-exposures. Using the measured value we estimate the false-alarm probability P($\chi^2 > \chi^2_{\mathrm{m}}$). RV variability parameter $\eta$ is then defined as $\eta = -\log{(1-P(\chi^2 > \chi^2_{\mathrm{m}}))}$, which is always positive, and a larger $\eta$ corresponds to a greater probability of the WD being in a binary. We use this parameter to select the binary candidates.

\begin{figure}
    \centering
    \includegraphics[width=\linewidth]{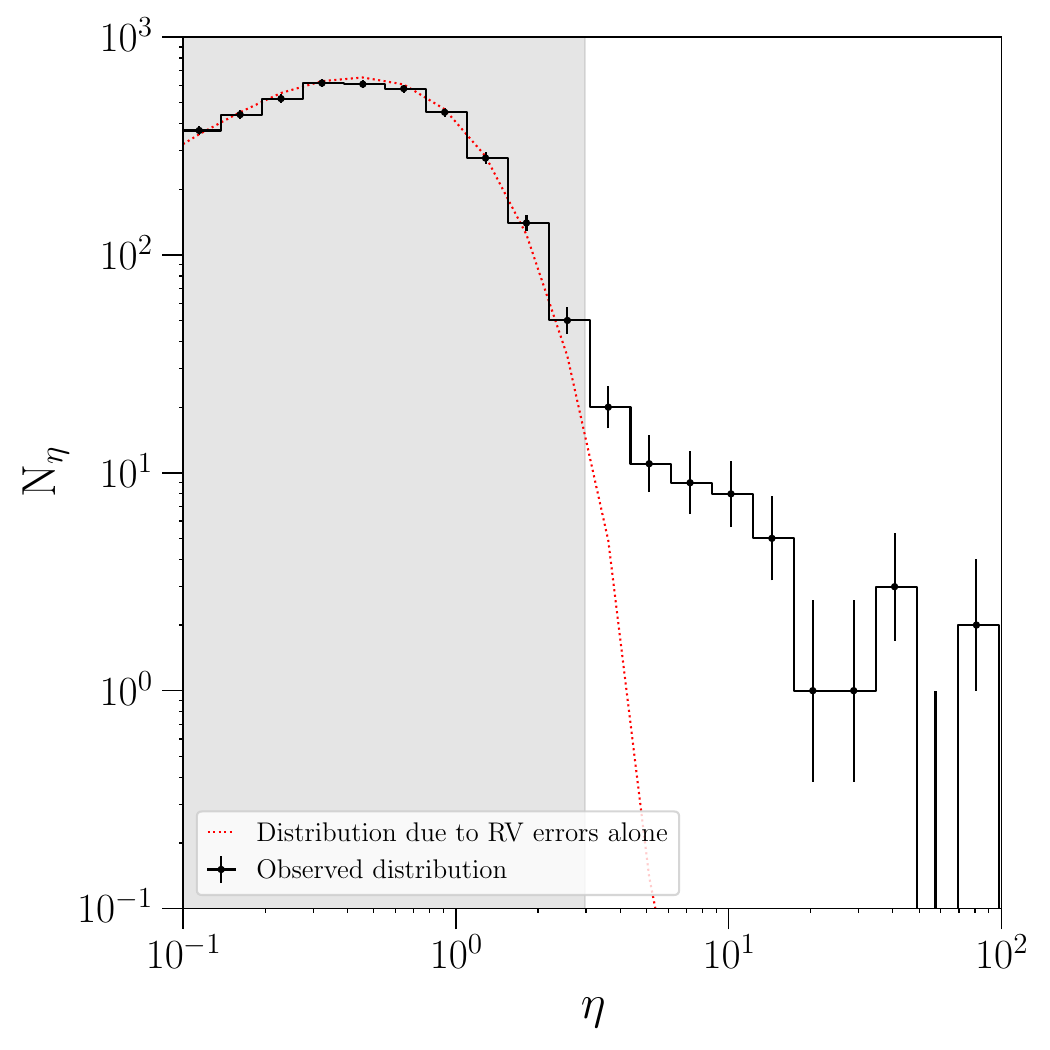}
    \caption{For different RV variability parameters $\eta$, the distribution of the number of WDs that show variability greater than $\eta$ is shown. In red we show the theoretical expectation, and in black we show the observed distribution. Our sample shows larger variability than expected, indicating the presence of binaries.}
    \label{fig:eta}
\end{figure}

\begin{figure*}
    \centering
    \includegraphics[width=\linewidth]{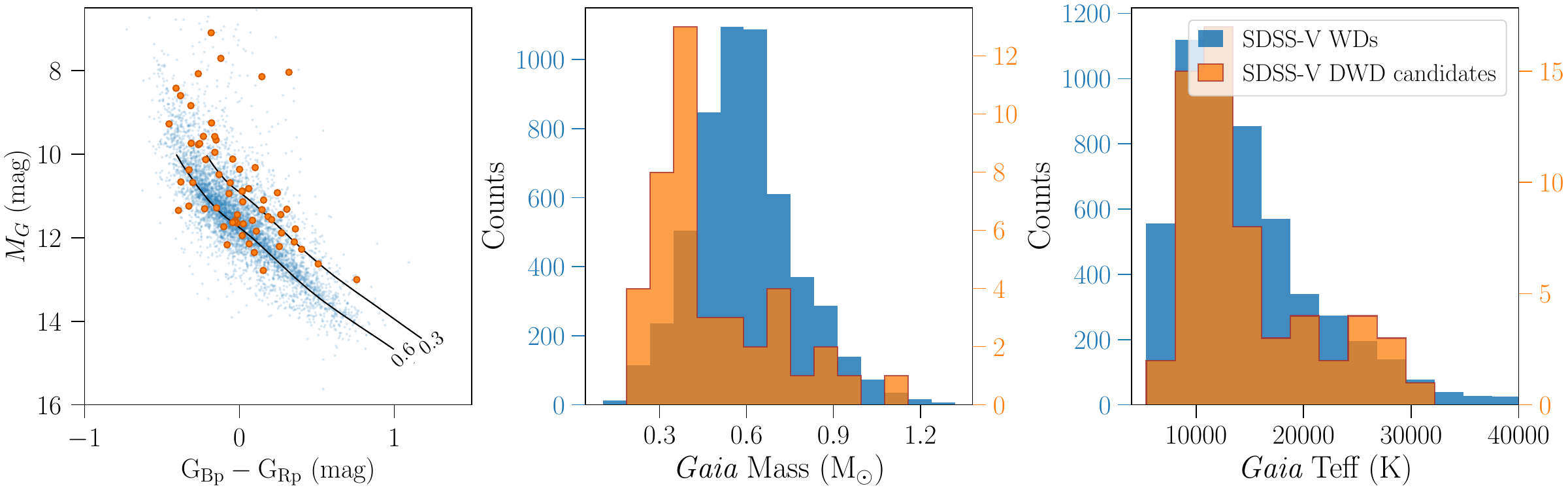}
    \caption{Left: The {\it Gaia} color-magnitude diagram of our parent sample of WDs with hydrogen lines (blue) and DWD binary candidates (orange). We also show the cooling curves of 0.3 M$_\odot$ and 0.6 M$_\odot$ WD \citep{bedard_spectral_2020}. Middle: The mass distribution of the parent sample and the DWD binary candidates. Right: The temperature distribution of the parent sample and the DWD binary candidates. We use the mass and temperature of the WDs obtained by \cite{gentilefusillo_gaia_2019} using \textit{Gaia} photometry}
    \label{fig:CMD_Distributions}
\end{figure*}

\section{Catalog of DWD binary candidates\label{sec:catalog}}
\subsection{RV Variables}

To estimate the value of $\eta$ that is most suitable to detect binaries, we simulate the SDSS-V observation pattern by randomly choosing the number epochs, cadence, and associated RV error of the observed WDs, and calculate the resulting $\eta$ distribution due to RV error alone. The result is shown in Fig.~\ref{fig:eta}. We find that $\eta > 3$ is a good cutoff beyond which the number of false-positive binaries sharply drops. Choosing a higher cut-off increases the likelihood of each candidate being a real binary, at the expense of the number of candidates. Decreasing the cut-off increases the number of binary candidates, but at the expense of increasing false positives. Choosing the right cut-off becomes important by balancing the number of candidates and false positives. In Sec.~\ref{sec:binary_population}, we discuss in detail how the theoretical distribution due to SDSS errors alone is generated.
 
We find 69 systems which satisfy the criterion $\eta > 3$. We visually inspect the spectra of these candidates with maximum and minimum RV, and discard six systems in which majority of exposures were affected by cosmic ray/spurious emission line features making the measured RVs unreliable. Three of these show emission features exactly at hydrogen Balmer lines and are cataclysmic variable (CV) candidates, which are briefly discussed in Appendix~\ref{appendeix:CV}. We list the remaining 63 DWD binary candidates in Table~\ref{tab:dwd_candidates}. Due to the distribution of RV errors shown in Fig.~\ref{fig:eta}, we estimate that about five WDs (mostly on the low $\eta$ end) are false positives giving us a sample of about 58 real DWD binaries. 

In Fig.~\ref{fig:CMD_Distributions} we show the color-magnitude diagram, the mass distribution, and the temperature distribution for both the binary candidates and rest of the SDSS-V WD sample. The binary population has preferentially lower photometric masses, as expected, since the low-mass WDs are larger and hence typically the visible star in DWD binaries. Additionally, binary population synthesis and observations suggest that lower-massed WDs are more likely to be found in binaries, with extremely low mass (ELM) WDs being found exclusively in binaries \citep{brown_binary_2011,toonen_supernova_2012,li_formation_2019,munday_dbl_2024}. The temperature distribution of DWD candidates closely follows the distribution of single WDs. 

We cross-match our sample with literature using SIMBAD, enabled by \texttt{astroquery}, to look for previously published binaries or binary candidates \citep{wenger_simbad_2000}. The results of the cross-match are presented in Table~\ref{tab:confirmed}. Nine DWD binary candidates have been confirmed to be binaries in the literature, and further two systems have been published as DWD binary candidates with SDSS DR14 data. There are three confirmed WD+main-sequence (MS) star binaries, six sub-dwarf candidates, which are typically found in binaries \citep{heber_hot_2016}, and may be precursors to extremely low mass (ELM) WDs and pre-ELM WDs \citep{pelisoli_sda_2018,adamane_pallathadka_discovery_2024}. In total, we find 30\% of the sample, 19 systems, to have existing evidence of binarity, and we present 43 new DWD binary candidates.

\begin{table*}
\centering
\begin{tabular}{ccccc}
$\eta$ & GAIA SOURCE ID & JNAME & Notes & Reference \\  \hline
 34.57 & 1319676603468508544 & J155708.48+282336.0 & WD+WD binary & \cite{brown_elm_2013} \\
 21.47 & 649261066445946624 & J082511.90+115236.4 & WD+WD binary & \cite{kilic_elm_2012}\\
 15.36 & 1500004000845782912 & J133725.22+395238.8 & WD+WD binary & \cite{chandra_99-minute_2021}\\
 13.21 & 3669445716389952768 & J142002.93+043903.5 & WD+WD binary candidate & \cite{yan_search_2024}\\
 12.54 & 3868927607051816320	& J103907.38+081841.0 & WD+WD binary & \cite{brown_binary_2011}\\
 10.91 & 1633145818062780544 & J174140.49+652638.7 & WD+WD binary & \cite{kilic_elm_2012}\\
 10.70 & 577257520277310848 & J090618.44+022311.6 & WD+WD binary & \cite{adamane_pallathadka_double_2025} \\
 9.46 &3168905043690276992 & J073616.22+162256.2 & WD+WD binary & \cite{breedt_using_2017}\\
 7.85 & 784186708135977088 & J112319.65+445045.6 & WD+WD binary candidate & \cite{yan_search_2024} \\
 3.06 & 2260805780286092032 & J180115.37+721848.7 & WD+WD binary & \cite{munday_dbl_2024}\\
 3.04 & 879036662822920448 & J074852.96+302543.4 & WD+WD binary & \cite{dobbie_two_2012} \\ 
 & & & & \cite{heintz_testing_2022} \\
 
 30.06 & 578539413395848704 & J085746.18+034255.3 & WD-MS binary & \cite{parsons_shortest_2012} \\
 12.57 & 1289020673097509760 & J150506.17+325959.4 & WD-MS binary & \cite{silvestri_new_2007} \\
 11.16 & 1609250784690803584	& J140357.66+541856.5 & WD-MS binary & \cite{silvestri_catalog_2006} \\
 & & & & \cite{rebassa-mansergas_post-common_2010} \\
 79.06 & 1017136594580182400 & J150506.17+325959.4 & sub-dwarf / pre-ELM WD candidate & \cite{pelisoli_sda_2018} \\
 37.75 & 3076962575704962176	& J083107.92+001331.3 & sub-dwarf / pre-ELM WD candidate & \cite{geier_population_2019} \\
 10.78 & 677695609668436736 & J081715.40+235141.6 & sub-dwarf / pre-ELM WD candidate & \cite{geier_population_2017} \\
 5.61 & 3214584494782707456 & J050919.88-022639.9 & sub-dwarf / pre-ELM WD candidate & \cite{culpan_population_2022} \\
 3.43 & 3095806242204017152 & J080645.61+053205.2 & sub-dwarf / pre-ELM WD candidate & \cite{geier_population_2019} \\
 3.24 & 3213132825902773376 & J045835.79-033051.2 & sub-dwarf / pre-ELM WD candidate & \cite{culpan_population_2022} \\
\\
 
\end{tabular}
\caption{The cross-match of our binary candidates with the literature}
\label{tab:confirmed}
\end{table*}

\subsection{Periodic Variables}
\begin{figure*}[t]
    \centering
    \includegraphics[width=\linewidth]{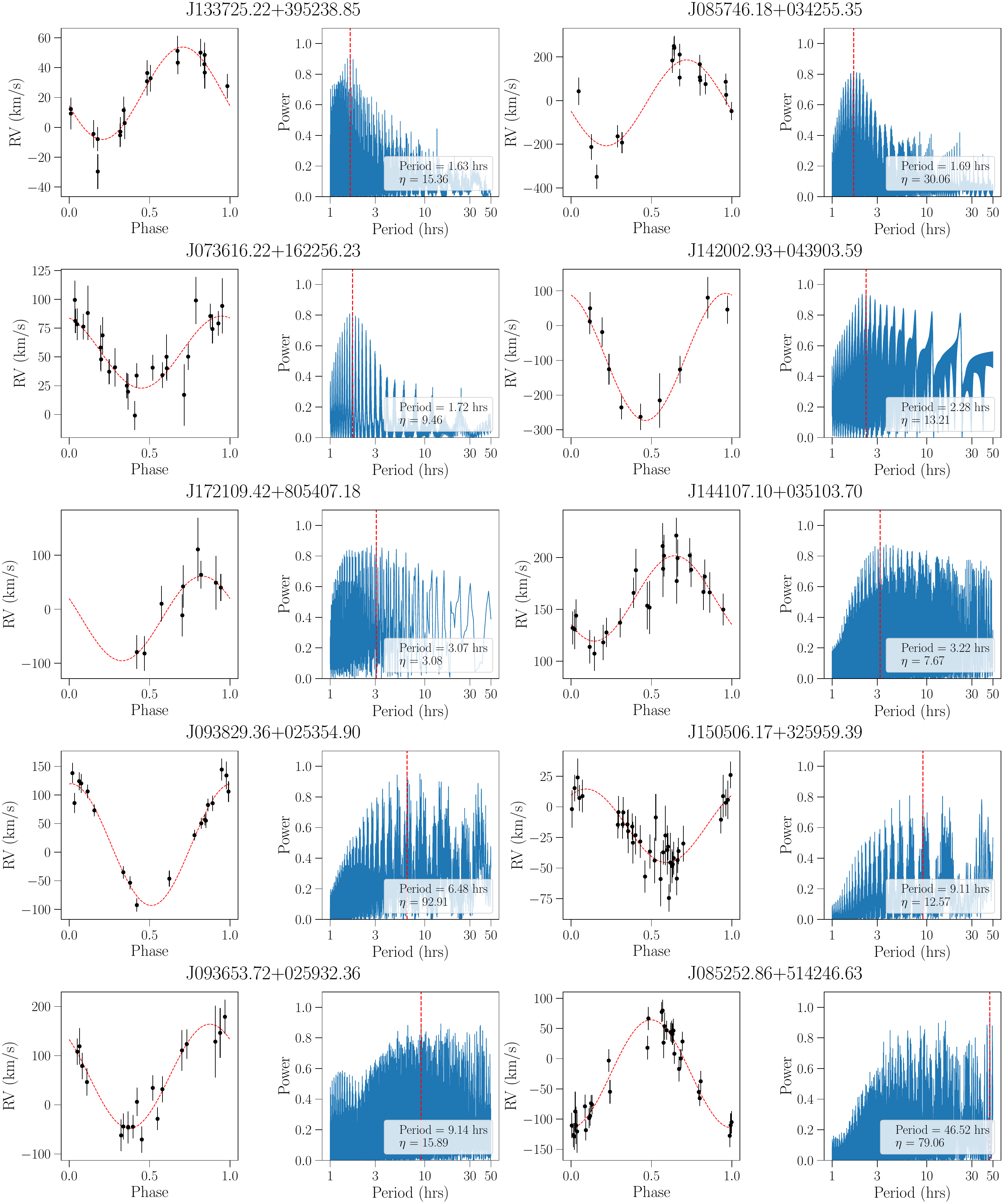}
    \caption{The results from the Lomb-Scargle periodogram period search for 10 WDs with large phase coverage in the RV curve. We show the best-fitting RV curve, along with the periodogram power. The best-fit period is chosen where the power is maximum.}
    \label{fig:periodogram}
\end{figure*}

We carry out a period search for all binary candidates using Lomb-Scargle periodogram \citep{lomb_least-squares_1976,scargle_studies_1982}. With the RV measurements, RV uncertainties, and observation times, we obtain the Lomb-Scargle periodogram through \texttt{astropy}, and the best-fit orbital period is chosen where the power is maximized. We looked through the best-fit results individually and found ten candidates which had RV curves with large phase coverage and high power in the power spectrum, which we show in Fig.~\ref{fig:periodogram}. The fit parameters are described in Table~\ref{tab:LS}. 

Three systems have periods measured in the literature using high-resolution follow-up observations : J133725.22+395238.85 is a double-lined DWD binary with a period of 1.6508 hours \citep{chandra_99-minute_2021}, J085746.18+034255.35 is a WD+MS binary with a period of 1.5623 hours \citep{parsons_shortest_2012}, and J073616.22+162256.23 is a DWD binary with 1.656 hours orbital period \cite{breedt_using_2017}. Using SDSS-V data alone, we derive a period of 1.63 hours for J133725.22+395238.85, 1.69 hours for J085746.18+034255.35, and 1.72 hours for J073616.22+162256.23, all of which agree with previously published values.
This comparison illustrates that even without follow-up data, in cases of good enough phase coverage by the SDSS data alone, we can obtain high-quality orbital periods. 


\begin{table}
\centering
\begin{tabular}{ccccc}
 JName & \multicolumn{2}{c}{Period} &$\eta$ \\ 
 & \multicolumn{2}{c}{(hrs)} & \\ 
 & This work & Published value &\\ \hline
 J133725.22+395238.8$^{\dagger}$ & 1.63 & 1.6508 & 15.36\\
 J085746.18+034255.3$^{\ast}$ & 1.69 & 1.5623 & 30.06\\
 J073616.22+162256.2 & 1.72 & 1.656 & 9.46\\
 J142002.93+043903.5 & 2.28 & -- & 13.21\\
 J172109.42+805407.1 & 3.07 & -- & 3.08\\
 J144107.10+035103.7 & 3.22 & -- & 7.67\\
 J093829.36+025354.9 & 6.48 & -- & 92.91\\
 J150506.17+325959.3 & 9.11 & -- & 12.57\\
 J093653.72+025932.3 & 9.14 & -- & 15.89\\
 J085252.86+514246.6 & 46.52 & -- & 79.06\\
\end{tabular}
\caption{The result of the Lomb-Scargle peridogram period search, along with the best-fit RV semi-amplitude. We also present the previously published values, whenever available. \newline $^{\ast}$J085746.18+034255.35 is a WD+MS binary with faint emission features. It is included in the table for completeness.}
\label{tab:LS}
\end{table}

\section{Constraints on DWD Binary population}
\label{sec:binary_population}
\subsection{Analysis}
We follow the prescription outlined by \citep{badenes_merger_2012,maoz_characterizing_2012,maoz_binary_2017,maoz_separation_2018} to constrain the binary fraction, and $\alpha$, the power law index of binary separation distribution at the formation of DWDs (assuming the distribution of binary separations, $n(a)\, \mathrm{d}a$, follows a power law of the form $n(a) \propto a^\alpha$). 
This involves comparing observations to simulation of mock SDSS observations for different binary population models.

We first remove the three confirmed WD+MS binaries from our sample. In this work, we select DWD binary candidates with $\eta > 3$. In principle, this can identify even wide binaries with relatively small RV variation, but makes it sensitive to WD binaries with invisible brown dwarf and M-dwarf companions. A visible M-dwarf companion would be identified in the SnowWhite pipeline, but to exclude any binaries with an invisible brown dwarf or M-dwarf companion with a mass less than 0.1 M$_{\odot}$, we select only those systems that show $\Delta \mathrm{RV_{max}} > 100$ km s$^{-1}$, where $\Delta \mathrm{RV_{max}}$ is the maximum RV shift observed across any two exposures of the same WD. The 100 km s$^{-1}$ constraint also sets a limit on the maximum orbital separation of binaries we can probe. A extreme mass-ratio binary of 0.2~M$_{\odot}$ WD and 1.2~M$_{\odot}$ WD separated by 0.4~AU can produce, at most, peak-to-peak variation of 100 km s$^{-1}$, and thus is approximately the maximum separation of binaries in our analysis.

To simulate the observations, we first build a mock population of WDs. The mass of the primary WD is drawn from the mass distribution of 11\,129 DA WDs with SNR$>$10 and $\mathrm{T_{eff} > 6000 K}$ from SDSS DR14 sample by \cite{kepler_white_2019}. 
The mass measurements of SDSS-V DA WDs by \citet{crumpler_large_2025} were made using photometric data and has larger uncertainties, while the \cite{kepler_white_2019} mass measurements used the spectroscopic data. Further, \cite{kepler_white_2019} mass distribution acts as useful tool for comparison with results from \cite{badenes_merger_2012} and \cite{maoz_binary_2017}. The primary mass is drawn from this distribution in the range of 0.2$-$1.2 M$\odot$. 

Following \citet{maoz_binary_2017}, when the primary mass is below 0.25 M$_{\odot}$ the WD is always assigned to be in a binary, and when the mass is in the range of 0.25 M$_{\odot}$$-$0.45 M$_{\odot}$ we assign a probability of 70\% for it to be in a binary \citep{brown_binary_2011}.
For all WDs above 0.45 M$_{\odot}$, we assign a probability of $f_{\mathrm{bin},0.4}$ to be in a binary with separation $<$0.4 AU. For the assumed distribution, we find that 10.4\% of all WDs to be a low-mass WD with mass less than 0.45 M$_{\odot}$, and 7.5\% to be a low-mass WD in a binary. This is much higher than \cite{badenes_merger_2012} and \cite{maoz_binary_2017}, who found $<$ 2\% of WDs to be a low-mass WD in a binary. This is primarily because we consider all WDs with $\mathrm{T_{eff}}$ above 6\,000 K. In contrast, if we consider only hot WDs ($>$ 12\,000 K) similar to \cite{maoz_binary_2017}, we find that only 2.8\% are low-mass WDs in binaries, comparable to \cite{maoz_binary_2017}. 

When the system is in a binary and the primary mass is above 0.45 M$_{\odot}$, the mass ratio is drawn from the distribution given by 
\begin{equation}
    P(q) \propto q^{\beta}, \quad q=\frac{m_2}{m_1},
\end{equation}
for 0.45 $\mathrm{M}_{\odot} < m_2 < m_1$. For $m_1 <$ 0.45 $\mathrm{M}_\odot$, $m_2$ is drawn from a uniform distribution between 0.2 M$\odot$ and 1.2 M$\odot$. As described in \cite{maoz_characterizing_2012}, these simulations are not very sensitive to mass ratio and we set $\beta = 0$ throughout.

DWD binary formation from main-sequence star binaries involves multiple stages of mass-transfer \citep{nelemans_reconstructing_2000,nelemans_population_2001,li_formation_2019,ivanova_common_2020}. These involve complex processes that are still not well understood and difficult to model. As discussed in \cite{maoz_characterizing_2012}, binary population synthesis simulations suggest that DWD binaries form with a separation distribution that follows a power law. Thus, on formation of a DWD binary, we assume the binary separation to follow a power law distribution with $n(a) \propto a^{\alpha}$. A fraction of extremely compact DWD binaries (period $<$ 10 minutes) can undergo a subsequent mass transfer stage leading to the formation of AM CVn binaries with periods in the range of 10-60 minutes \citep{nelemans_population_2001-1,solheim_am_2010}. However, these systems are typically characterized by strong emission lines and, in most cases, lack hydrogen in the spectra \citep{solheim_am_2010}, and are not expected to contaminate our sample.

Assuming a constant binary formation rate over the age of the Galaxy, and taking into the orbital decay due to gravitational wave emission, \cite{maoz_characterizing_2012} show that the current binary separation distribution is given by 

\begin{align}
    N(x) &\propto x^{4+\alpha}\left[\left(1 + x^{-4}\right)^{(\alpha+1)/4} - 1\right], \quad &\alpha \neq -1, \\
    N(x) &\propto x^3 \ln(1 + x^{-4}), \quad &\alpha = -1,
\end{align}

Here, $x = \frac{a}{(K t_0)^{1/4}}$, where $K = \frac{256}{5} \frac{G^3}{c^5} m_1 m_2 (m_1 + m_2)$, $t_0$ is the age of the Galaxy, $G$ is the gravitational constant, and $c$ is the speed of light. We sample this distribution between $a_{\mathrm{min}} = 2 \times 10^4$ km (at contact between the merging WDs) and $a_{\mathrm{max}} = 0.4$ AU.

When drawing the binary separation, binaries in certain ranges with low probabilities may not be sampled well and thus add to the Poisson random error. To account for this, we follow the prescription described by \cite{maoz_characterizing_2012}. We split the $a_{\mathrm{min}}-a_{\mathrm{max}}$ range into 10 different bins. We populate all bins uniformly and within each bin we draw from the binary separation distribution. We then assign a relative weight to each binary equal to the integral of the binary separation distribution over the bin it is in.

For the drawn masses and binary separation, we calculate the orbital period and the RV semi-amplitude assuming a circular orbit. When the orbital period is less than 15 minutes, any RV shift in the absorption lines is smeared out over the 15 minute SDSS exposure. For such binaries, we set the orbital velocity to zero. The distribution of WD temperatures and the inverted mass-radius relationship of WDs makes the choice of photometric primary unclear, unless the WD is a low-mass WD. We randomly choose one of the WDs to be the photometric primary, and if mass of one of the WDs is less than 0.35 M$_{\odot}$ we always assign it to be the photometric primary. We then assign an inclination to each binary and sample the SDSS-V observation pattern to set the number of exposures and cadence of observation, and perform mock RV measurement. For lone WDs we assign a zero RV. Finally, we assign an error to each binary by drawing from the observed RV error distribution and add normally distributed errors to the RV observations. We find a correlation between RV errors and the number of epochs observed per object, with WDs with a greater number of epochs having lower median RV errors. Hence, we draw the RV errors and the observation pattern simultaneously from the SDSS-V observations corresponding to the same WD.

We simulate a mock sample of WDs and the associated RV measurements using the above procedure. For both the observed WDs and the simulated WDs, we calculate $\eta$ and $\Delta \mathrm{RV_{max}}$, the maximum RV shift observed across any two exposures of the same WD. \citet{badenes_merger_2012} and \citet{maoz_binary_2017} directly compared the observed and simulated $\Delta \mathrm{RV_{max}}$ distribution. However, we find that the SDSS-V RV error distribution is broader than the RV error distribution in \citet{badenes_merger_2012}, which causes the binaries to be buried among other noisy observations making the direct comparison impossible. To remove the noisy observations, we first select all WDs with $\eta > 3$, and then compare the $\Delta \mathrm{RV_{max}}$ distribution. The simulated distribution is scaled by the ratio of the total number of WDs in the observed sample to the simulated sample, and we generate a histogram of the $\Delta \mathrm{RV_{max}}$ distribution with 50 km s$^{-1}$ bins. The model likelihood given the observed data can be calculated as the product of Poisson probabilities of finding the number of observed systems in each bin given the theoretical expectations for that bin, across all bins of interest greater than 100 km s$^{-1}$. We also compared the $\eta$ distribution which works well, but leads to more relaxed constraints.


We explore a parameter space of $f_{\mathrm{bin}}$ and $\alpha$ to compute the likelihood contours, and simulate 0.5 million WDs for each ($f_{\mathrm{bin}},\alpha)$ pair. In total, we run 450 million simulations. 

\subsection{Validation}

\begin{figure}
    \centering
    \includegraphics[width=\linewidth]{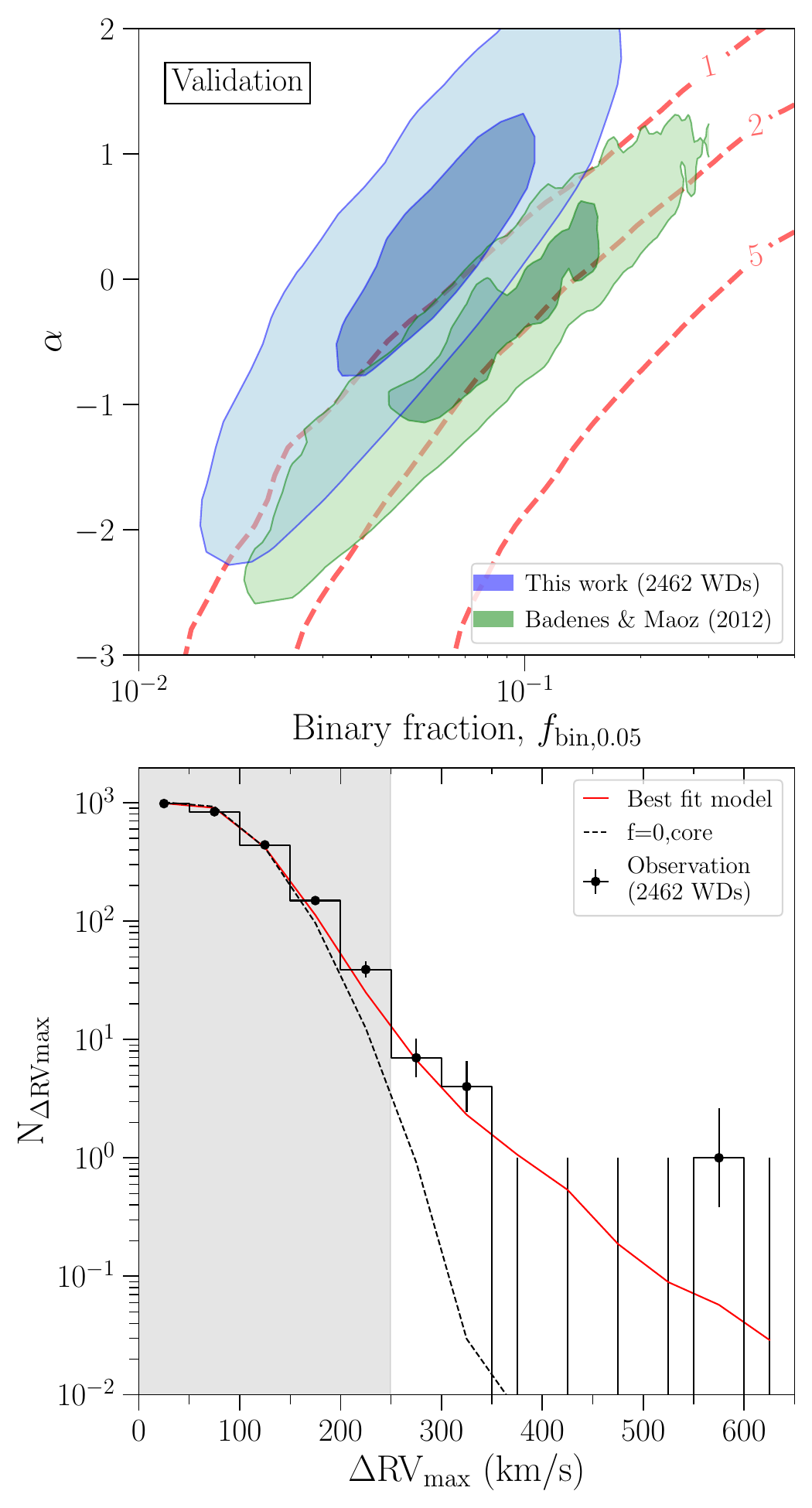}
    \caption{The best-fit 1-$\sigma$ (darker shade) and 2-$\sigma$ (lighter shade) contours produced using the validation sample ($G < $ 19 mag and $\mathrm{RV_{err}} <$ 55 km s$^{-1}$) is compared the contours from \cite{badenes_merger_2012}. Both contours are 1-$\sigma$ smoothed using a Gaussian filter. The red dashed lines represent the expected number of super-Chandrasekhar mass binaries that can merge within the age of the universe.}
    \label{fig:2012_compare}
\end{figure}

We validate our simulation by directly comparing our results achieved with SDSS-V data with \cite{badenes_merger_2012}. Although both works rely on SDSS observations, the primary difference is the observation pattern and target selection between different generations of SDSS data. To achieve a sample similar to \cite{badenes_merger_2012}, we considered only WDs with $G$ band magnitude $<$19 mags and exposures with $\mathrm{RV_{err}} <$ 55 km s$^{-1}$, which resulted in a RV error distribution that closely resembles that of \cite{badenes_merger_2012}. Similar to their analysis, this validation sample with narrower RV error distribution probes binaries within 0.05 AU making it ideal for direct comparison. For the validation, we do not apply any $\eta$ cuts and directly compare $\Delta\mathrm{RV_{max}} > 250$ km s$^{-1}$ part of the distribution, to match the analysis of \cite{badenes_merger_2012}. With these cuts, we obtain a sample of 2462 WDs with more than one exposure.  After applying the binarity flags and drawing the WD masses similar to \cite{badenes_merger_2012}, we obtain the final contours shown in Fig.~\ref{fig:2012_compare}. We find that our result is consistent with \cite{badenes_merger_2012} to within 2-$\sigma$, despite the smaller sample size and differing observation pattern, which validates our analysis.

\subsection{Results}

\begin{figure}
    \centering
    \includegraphics[width=1\linewidth]{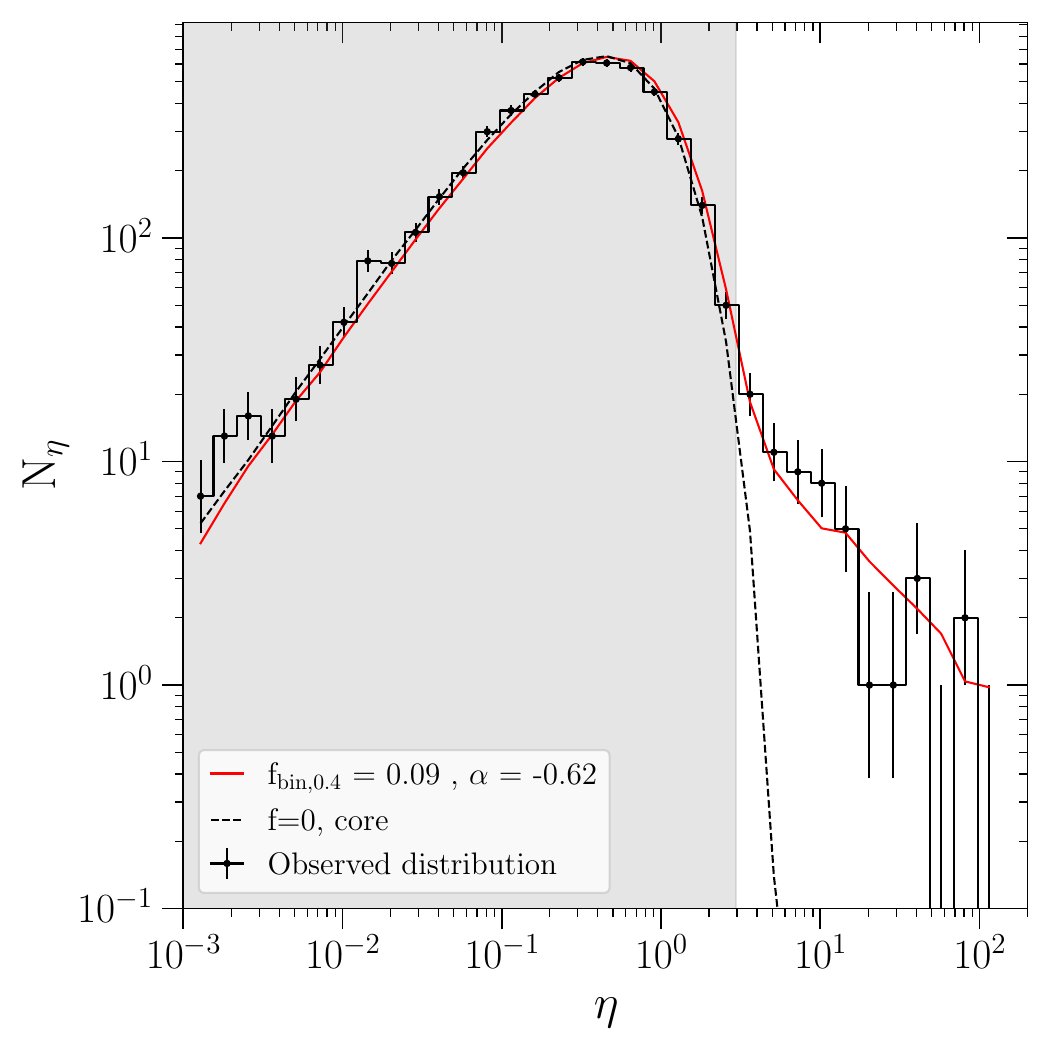}
    \includegraphics[width=1\linewidth]{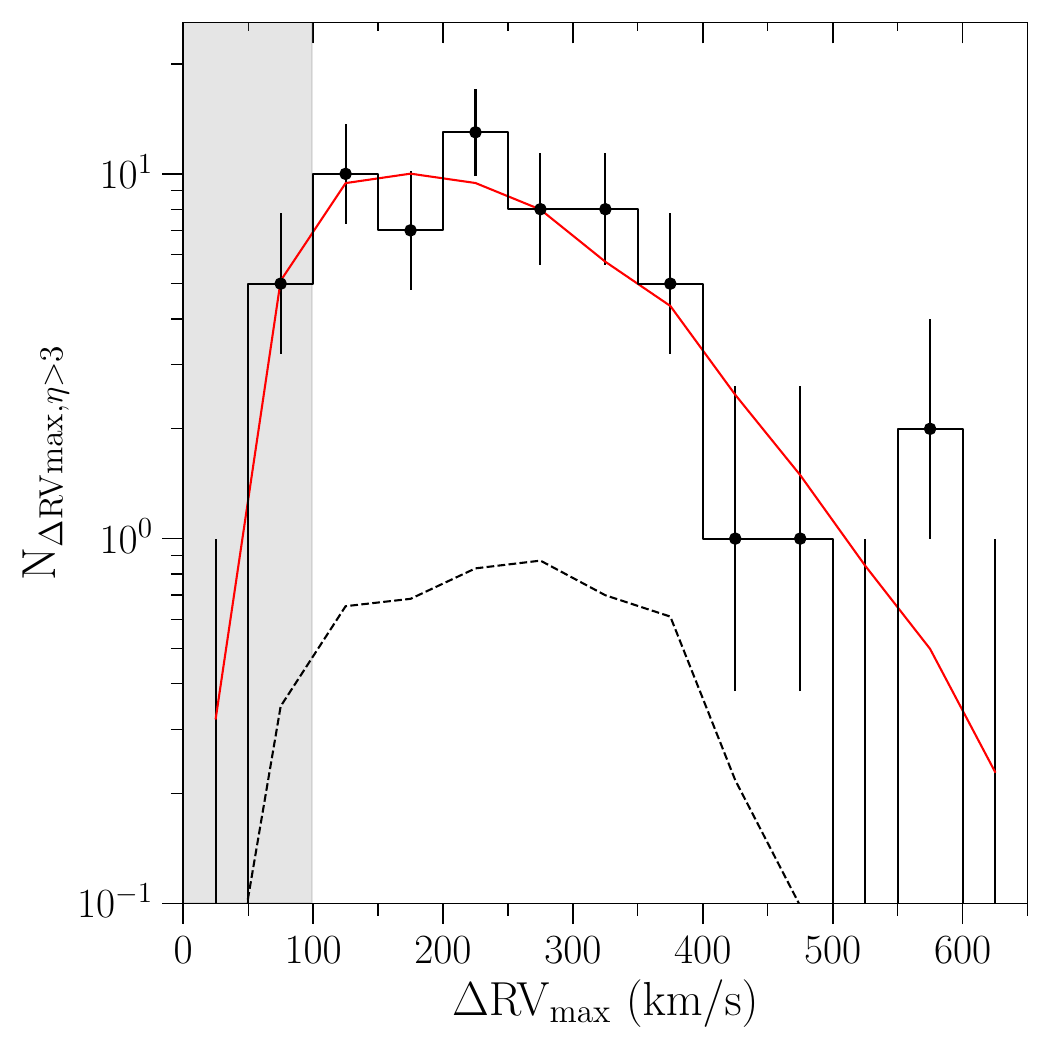}
    \caption{The results are presented for low-mass WD binary fraction of 7.5\%. Top: The distribution of $\eta$ is presented along with best-fit model. The greyed out region corresponds to $\eta<3$. The model agrees well with the data, despite the best-fit being calculated using the $\Delta\mathrm{RV_{max}}$ distribution. Bottom: $\Delta\mathrm{RV_{max}}$ distribution is presented for WDs with $\eta>3$. We find good agreement with the model even for data points not included in the fit.}
    \label{fig:bestFit_0.07}
\end{figure}

\begin{figure}
    \centering
    \includegraphics[width=\linewidth]{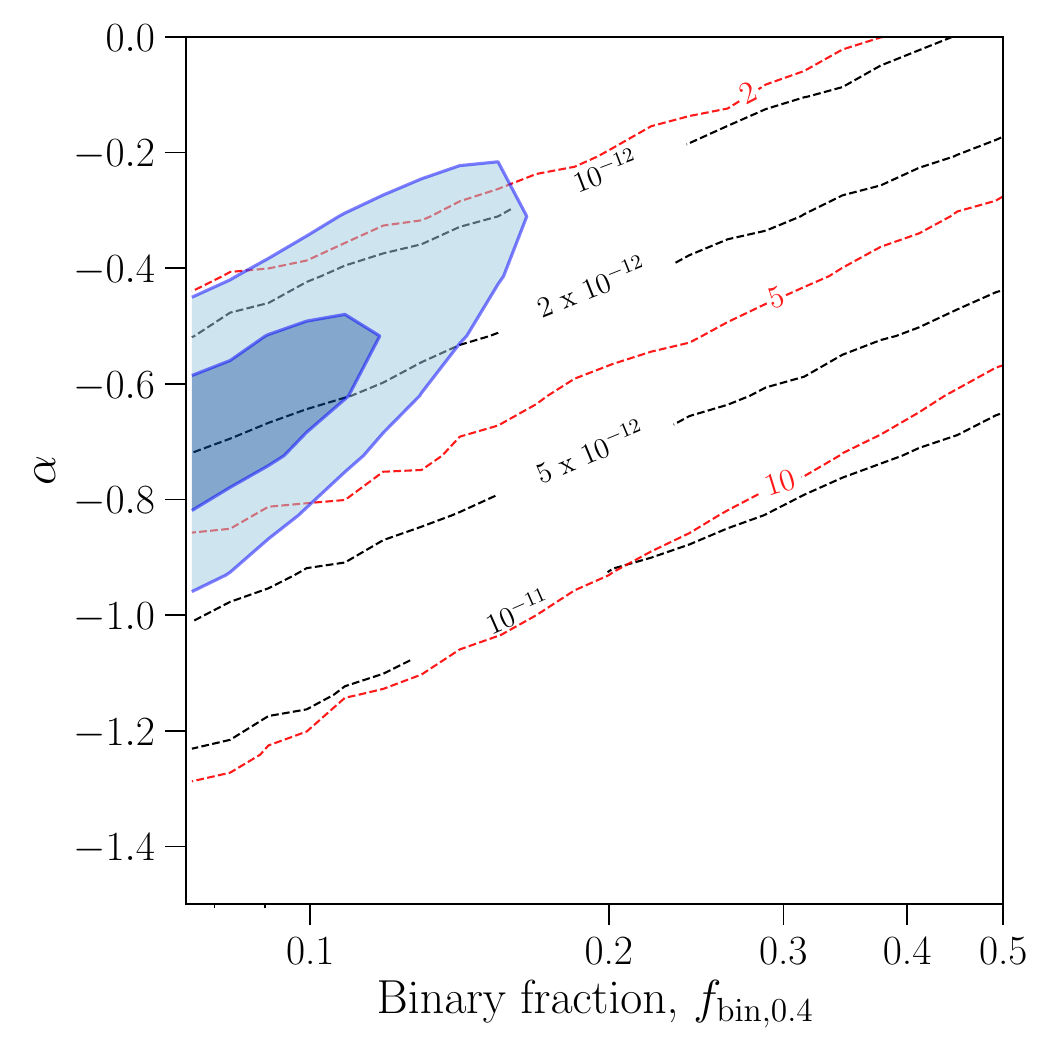}
    \caption{The results are presented for low-mass WD binary fraction of 7.5\%. The constraints on $f_{\mathrm{bin},0.4}$ and $\alpha$ are shown. The black dotted lines represent the DWD merger rate per year. The red dashed lines represent the number of super-Chandrasekhar mass binaries that can merge within the age of the universe.}
    \label{fig:simul_0.07}
\end{figure}

The final result of the fit is presented in Fig.~\ref{fig:bestFit_0.07} and the likelihood contours are shown in Fig.~\ref{fig:simul_0.07}. We find the best-fit values of $f_\mathrm{bin,0.4} = 0.09_{0.01}^{0.03}$ and $\alpha = -0.62_{0.10}^{0.10}$. There is a strict lower limit to binary fraction set by the low-mass WDs. We find a good agreement between the data and the model in both distributions presented in Fig.~\ref{fig:bestFit_0.07}, including the greyed out regions, although only $\Delta\mathrm{RV_{max}}>100$ km s$^{-1}$ points are fit. The fraction of low-mass WD binaries in our analysis is much higher compared to older studies because they considered $\mathrm{T_{eff}} > $12\,000 K, which excluded a large fraction of low-mass WDs. Hence, we do not place joint constraints and discuss this further in Sec.~\ref{sec:Discussion}. 

In Fig.~\ref{fig:simul_0.07}, in black dashed lines we show the DWD binary merger rate per year. Each simulated binary has a fixed merger time because of gravitational wave emission. We define the current merger rate as the number of binaries that can merge within the next million years divided by that time. We estimate the DWD merger rate to be $\approx 2 \times 10^{-12} \mathrm{yr^{-1}}$, a factor of five smaller than that found by \cite{maoz_separation_2018}. However, this difference is because of the differences in mass distributions used, and when this difference is taken into account our results are consistent with \cite{maoz_separation_2018}, as we show in Sec.~\ref{sec:con_bin_pop}.

\section{Discussion \label{sec:Discussion}}
\subsection{DWD binary sample}
In this paper, we present the catalog of 63 high confidence DWD binary candidates in SDSS-V DR19. We cross-matched our catalog with literature and find that 30\% of the sample has been confirmed either be confirmed binaries or highly likely to be RV variables.
This boosts confidence in our candidates. Follow-up observations of the rest of the candidates will uncover a large number DWD binaries. With our selection criterion of $\eta > 3$, we expect about five false positives, giving us a realistic DWD sample size of 58 out of 63 candidates. We compare our DWD candidate sample with the rest of our parent sample in Fig.~\ref{fig:CMD_Distributions}. We find that the the photometric primaries in binaries have lower photometric masses compared to rest of the WD sample, while the temperature distribution of DWD candidates closely follows the distribution of individual WDs. The most massive binary candidate in our sample is J074852.96+302543.4 with a \textit{Gaia} WD mass of 1.09 M$_{\odot}$. This is a DWD wide binary candidate with a magnetic and non-magnetic WD \citep{dobbie_two_2012,heintz_testing_2022}. The catalog also has WDs of rare types which are interesting for follow-up observations. J180115.37+721848.7 is a ZZ Ceti and is also confirmed to be a double lined DWD binary \citep{romero_discovery_2022,munday_dbl_2024}. J112328.49+095619.3 is a high confidence DWD candidate with a variability parameter of $\eta = 7.88$ and is classified as a magnetic WD \citep{kepler_magnetic_2013}. J023543.07+005557.1 is a candidate for WD with gaseous debris disk with weak calcium emission lines \citep{gansicke_sdssj104341530855582_2007}. J090618.44+022311.6 is only the fourth short period DWD binary with hydrogen atmosphere DA WD and carbon atmosphere DQ WD \citep{adamane_pallathadka_double_2025}.

\subsection{Constraints on the binary population}
\label{sec:con_bin_pop}
While our catalog is not exhaustive list of binaries in the SDSS-V sample, it provides a sizable set of high confidence binary candidates that can be used for DWD binary population studies. We used our catalog to place constraints on the binary fraction in the WD population and the binary separation distribution. We find best-fit values of $f_\mathrm{bin,0.4} = 0.09_{-0.01}^{+0.03}$ and $\alpha = -0.62_{+0.10}^{-0.10}$, and the DWD merger rate is $\approx 2 \times 10^{-12} \mathrm{yr^{-1}}$. 

Our measurements are most sensitive to the binary fraction and the separation distribution at $a_max<0.4$ AU, whereas those of \cite{badenes_merger_2012} and \cite{maoz_binary_2017} lie on $a_{\mathrm{max}} < 0.05$ AU and $a_{\mathrm{max}} < 4 $ AU, respectively. The fraction of low-mass WD binaries in our sample is higher than the previous works. To gauge the agreement between different constraints on the underlying binary population and to understand the effect of the mass distribution on the final result, we restrict the mass distribution to the sample of hot WDs ($\mathrm{T_{eff}>12\,000~K}$) as in \cite{maoz_binary_2017}, and redo the analysis. We obtain a lower limit of 2.8\% on the binary fraction from low-mass WDs. Following \cite{maoz_separation_2018},  we transform the new likelihood contours into a common parameter space with $a_{\mathrm{max}} < 4 $ AU by mapping likelihoods $L(f_{\mathrm{bin},a_\mathrm{max,1}},\alpha) \longrightarrow L(f_{\mathrm{bin},a_\mathrm{max,2}},\alpha)$
with 

\begin{equation}
    f_{\mathrm{bin},a_\mathrm{max,2}} = \frac{\int_{a_\mathrm{min}}^{a_\mathrm{max,2}}n(a,\alpha) \mathrm{d}a} {\int_{a_\mathrm{min}}^{a_\mathrm{max,1}}n(a,\alpha) \mathrm{d}a}f_{\mathrm{bin},a_\mathrm{max,1}}.
\end{equation}

\begin{figure}
    \centering
    \includegraphics[width=\linewidth]{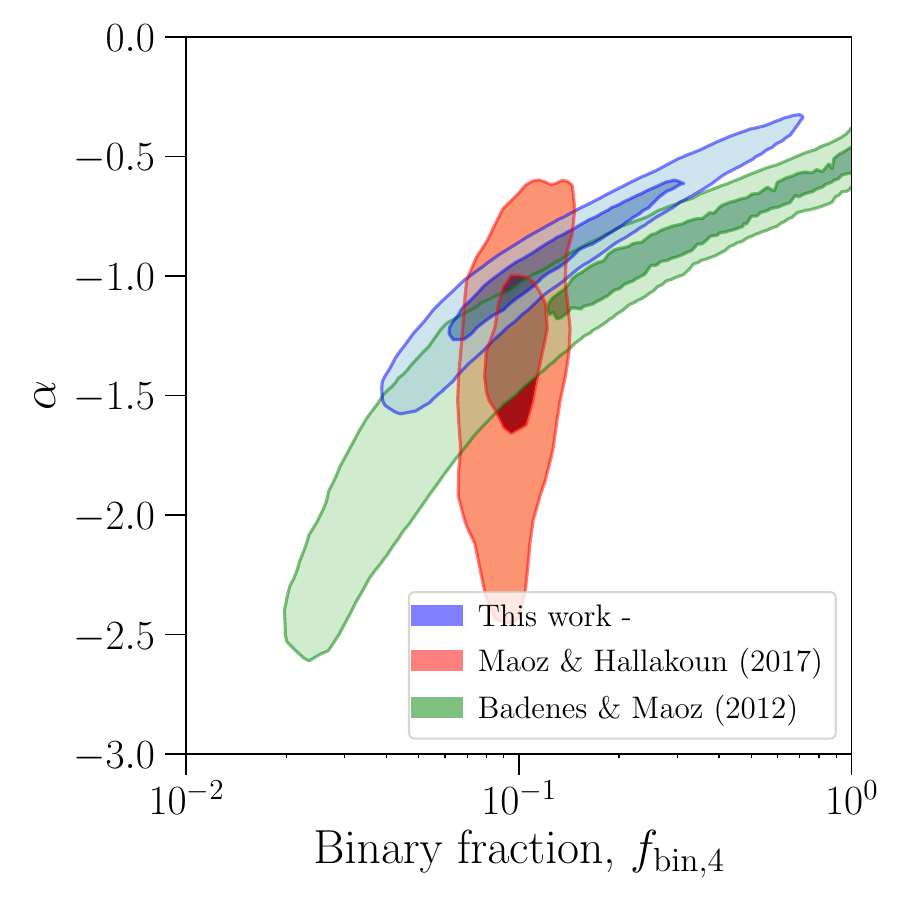}
    \caption{
    Likelihood contours from \cite{badenes_merger_2012}, \cite{maoz_binary_2017}, and this work (for hot WDs), all for binaries with separation less than 4 AU. The results of \cite{badenes_merger_2012} after transformation to $< 4$ AU space visually disagrees with those presented in \cite{maoz_separation_2018}. The choice of smoothing and how to discard points that get transformed to $f_{\mathrm{bin}} > 1$ values can significantly alter the contours. Here, we perform 1-$\sigma$ smoothing of the contours before the transformation, and then project the transformed contours on to a dense grid with $f_{\mathrm{bin}} < 1$.}
    \label{fig:combined_contours}
\end{figure}

Points that end up in $f_{\mathrm{bin}} > 1$ after transformation are discarded by interpolating on to $f_{\mathrm{bin}} < 1$ grid.  The results are shown in Fig.~\ref{fig:combined_contours}. When transformed, our results for hot WDs agree well with the previously published results, and they all overlap. We find that the $1\sigma$ and $2\sigma$ contours are well constrained and are shifted to lower values of $\alpha$ and of the binary fraction $f$ compared to Fig.~\ref{fig:simul_0.07}. This is primarily because a low-mass WD can exhibit higher RV variation and behaving similarly to a more massive WD but in short period binary. Thus, for the same binary fraction, the population with higher low-mass WDs fraction would need to be in wider orbits, compared to more massive WDs, to exhibit a similar RV variation distribution, and thus shifting the contours to higher $\alpha$. 

\subsection{Where are the Type-Ia Supernovae progenitors?}
\begin{figure}
    \centering
    \includegraphics[width=\linewidth]{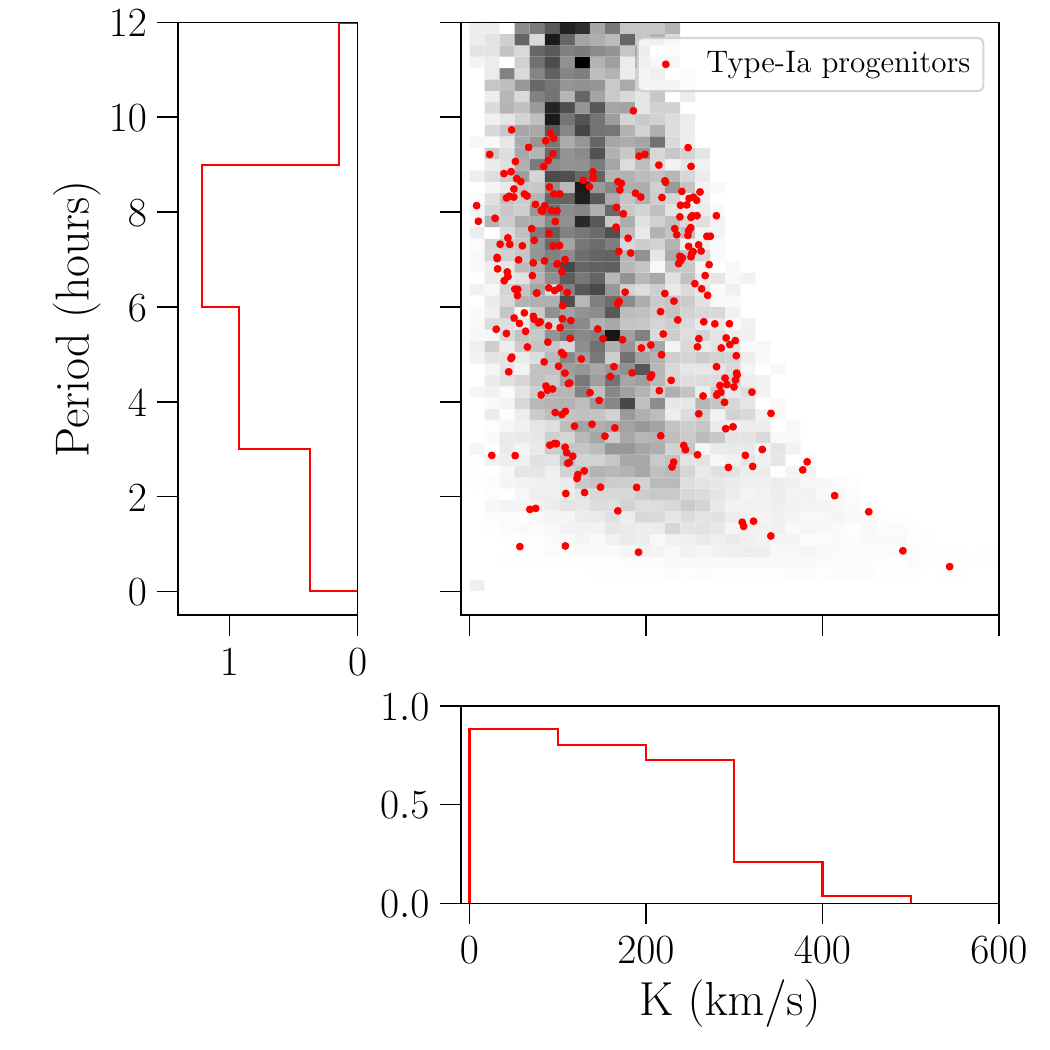}

    \caption{Predicted period and RV semi-amplitude for the Type-Ia progenitors (red) and for all DWDs (grey) in the simulated SDSS-V WD sample. The number of Type-Ia progenitors is exaggerated in the central plot to demonstrate the distribution of the sample. The histograms in bottom and left plot show the distribution of RV semi-amplitude K and the period P for Type-Ia progenitors. We find that a Type-Ia progenitors are concentrated towards longer periods and thus relatively modest RV semi-amplitudes, making them difficult to identify.}
    \label{fig:type_ia}
\end{figure}

For each simulated sample, we estimate the number of super-Chandrasekhar mass binaries that can merge within the Hubble time. For the best-fit model in Fig.~\ref{fig:simul_0.07}, we expect about 4 such binary systems, with 3$\sigma$ range extending between 0$-$10 . In Fig.~\ref{fig:type_ia} we show the distribution of period and the RV semi-amplitude for the DWD sample with Type-Ia progenitors highlighted in red. The number of highlighted Type-Ia progenitors are exaggerated to demonstrate the distribution of progenitors while the absolute numbers are preserved in the histograms. The period distribution poses the biggest problem in the discovery of Type-Ia supernova progenitors. With a typical SDSS pattern of three consecutive exposures, a binary with a period of 9 hours and 300 km s$^{-1}$ RV semi-amplitude, would show a peak-to-peak RV variation of an unremarkable 100 km s$^{-1}$. Even with several exposures, a periodic variation greater than about 7 hours is difficult to probe without a high-precision instrument. This is seen in our binary catalog Fig.~\ref{fig:periodogram} : binary candidates J150506.17+325959.3, J093653.72+025932.3, and J085252.86+514246.6 have tentative periods greater than 9 hours but, despite the availability of a large number of observations, it cannot be conclusively determined whether these systems have periods less than 10 hours or more.
Small follow-up programs to ascertain the period range and mass range can be valuable before large scale follow-up efforts to determine the complete orbital solutions. Further, simultaneous, spectroscopic and photometric analysis can identify overluminous DWD binaries where photosphere of both stars become visible, which can be particularly valuable to follow-up and identify super-Chandrasekhar mass binaries before detailed period analysis \citep{chandra_99-minute_2021,munday_dbl_2024,munday_super-chandrasekhar_2025}.

While the simulations are insensitive to small changes in $\beta$, the power law index of the mass ratio distribution, significant changes in the fraction of low-mass WDs in the WD sample can alter the results of the simulation. This problem is also tightly connected to the problem of finding super-Chandrasekhar mass binary candidates in our sample and of calculating the DWD merger rate based on simulations. In Fig.~\ref{fig:2012_compare}, we show the predicted number of Type-Ia progenitors in red dashed lines in the validation sample. This result follows the binarity flag of \cite{badenes_merger_2012} with extremely low-mass WDs ($<$ 0.25 M$_{\odot}$) always assigned to be in binaries, while the rest of WDs have probability $f_{\mathrm{bin,0.05}}$ to be in binaries. This results in 0.07\% of WDs to be low-massed and nearly all binaries are distributed throughout the WD mass range. Thus, super-Chandrasekhar mass binaries can only occur when $m_1$ is drawn from the high mass end of WD mass distribution. Only about 1.5\% of WDs have mass greater than 1 M$_{\odot}$ and at $f_{\mathrm{bin,0.05}} = 0.1$, this results in only 0.15\% of WDs being found in a binary with a mass of at least 1 M$_{\odot}$. In contrast, the result in Fig.~\ref{fig:simul_0.07} assumes that 70\% of WDs with mass less than 0.45 M$_{\odot}$ are found in binaries, and about 10\% WDs have mass less than 0.45 M$_{\odot}$, leading to 7\% of all WDs being found in low-mass WD binaries. For each of those binaries, the secondary mass is drawn uniformly between 0.2-1.2 M$_{\odot}$, leading to a 20\% chance that the secondary mass is greater than 1 M$_{\odot}$. This results in 1.4\% of all WDs being found in binaries with at least one WD having a mass greater than 1 M$_{\odot}$. 
These large differences in the composition of the binary population based on different assumptions highlights the need for accurate mass measurements of the underlying WD population. 

Another aspect that can impact the conclusions about Type-Ia progenitors is the magnitude limited nature of SDSS survey. Magnitude limited surveys are affected by selection cuts and the low-mass WDs, which are larger and typically more brighter than high mass WDs, are overrepresented. The volume limited 100 pc sample by \cite{kilic_100_2020} estimates the low-mass WD fraction for WDs with $\mathrm{T_{eff}} >$ 6\,000 K to be 4.5\%. This results in low-mass WD binary fraction of 3.2\%, and we estimate that the final constraints should be similar to Fig.~\ref{fig:combined_contours} with a similar fraction of low-mass WD binaries. Future works towards analyzing these binary samples by taking into account the magnitude limited nature of these surveys can provide more robust constraints and a clearer picture of DWD population.

\subsection{LISA detectable sources}

Laser Interferometer Space Antenna (LISA) is a space based gravitational wave detector currently in development that has the potential to detect a large number of Galactic compact binaries in the millihertz range (binary orbital period less than a few hours). In our catalog there is one confirmed LISA detectable binary: J133725.22+395238.8 is a double lined DWD binary with 99 minute orbital period \citep{chandra_99-minute_2021} and is expected to detected by LISA in the first four years of its run. In addition to potentially discovering the Type-Ia supernovae progenitors, LISA will detect DWD binaries that will be valuable for understanding the binary dynamics including complex processes involved in binary evolution  \citep{nelemans_gravitational_2001,ruiter_lisa_2010}. 

\begin{figure}
    \centering
    \includegraphics[width=\linewidth]{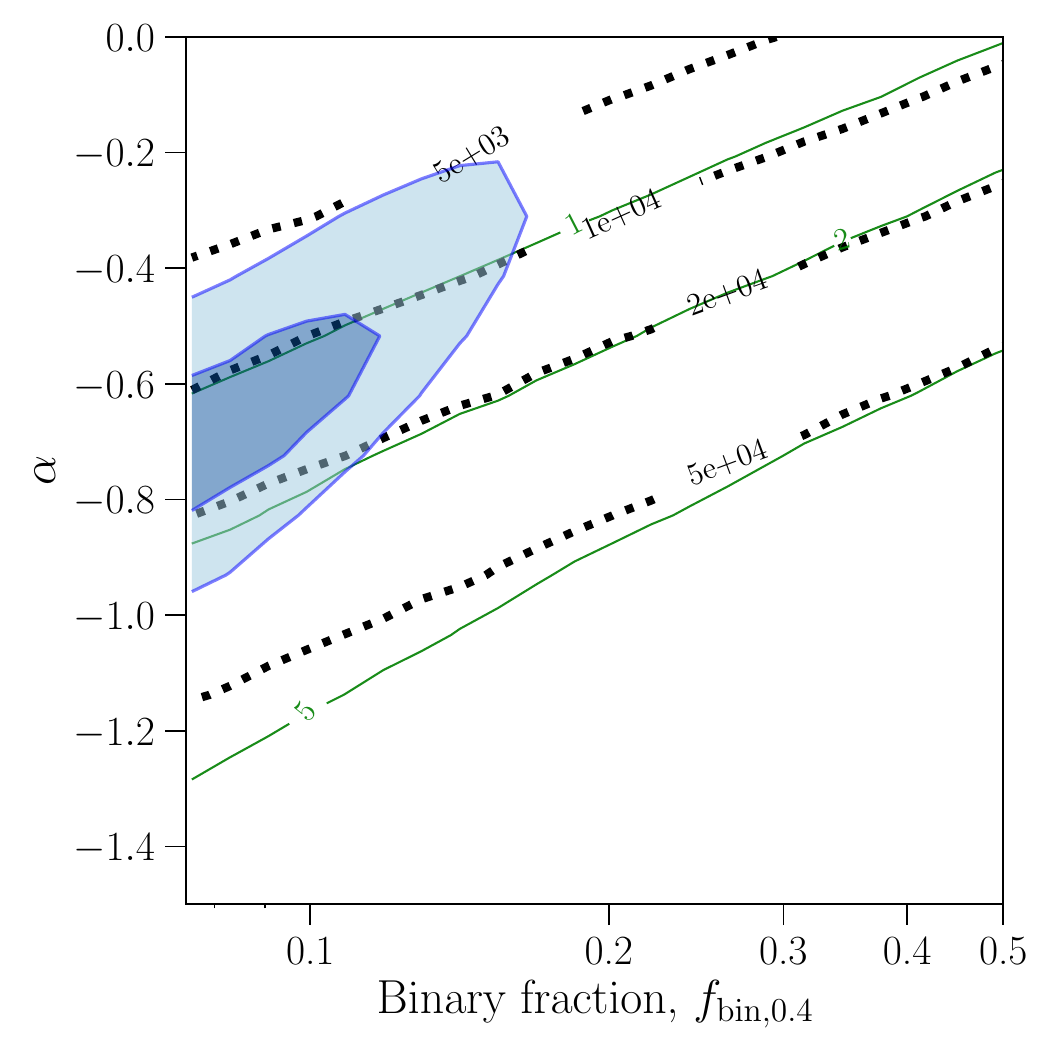}
    \caption{We show the expected number LISA detectable binaries in the first four year of its run. In green solid lines we show the predicted number of such binaries in our sample, and in black dotted lines we show the predicted number of such binaries in the galaxy.}
    \label{fig:lisa}
\end{figure}

Using the simulated sample, we estimate the expected number of LISA detectable sources in the observed SDSS-V sample. Each simulated binary has a maximum distance up to which the characteristic strain is greater than the strain detectable by LISA during the first 4 years of its run and is detectable as a gravitational wave source. We calculate this distance using the formulation for characteristic strain given by \cite{kupfer_lisa_2018} and the 4-year sensitivity curve given by \cite{robson_construction_2019}. We then assign a weight to such a system, equal to the probability of it being observed by SDSS-V at a distance less than the calculated maximal distance using the distance distribution of SDSS-V WDs. Finally, we calculate the sum of the weights to obtain the expected number of LISA sources in our sample. The result is shown as green solid lines in Fig.~\ref{fig:lisa}. We find that at 2$\sigma$ level, we expect around 2 such gravitational wave sources in our sample.

We extend this analysis further and calculate the expected number of DWD binaries detectable as LISA sources in the Galaxy. We approximate Milky Way as a thin disk with 300 pc scale height and 20 kpc radius. \cite{giammichele_know_2012} estimate the space density of WDs to be $4.39 \times 10^{-3}$ pc$^{-3}$. This gives us a total number of 1.7 billion WDs in Milky Way, and we scale our simulations to this number. For each simulated binary we again calculate the maximum distance up to which it is detectable as a LISA source. We then calculate the probability of such a binary falling in the thin disk of Milky Way within the calculated maximum distance from Earth, and assign it as a weight to the binary. The sum of weights over all such binaries gives the expected number of LISA sources in the galaxy. 

The result is shown in Fig.~\ref{fig:lisa} as thick dotted black lines. We find that the expected number of DWD binary LISA sources is in the range of 10\,000 -- 20\,000 and is within 50\,000 at 3$\sigma$ level. Our result is only a rough estimate and does not take into account the expected signal-to-noise ratio of such a source in LISA. But the predicted number is in agreement with results that rely on binary population synthesis models by \citet{nelemans_gravitational_2001} and \citet{lamberts_predicting_2019} -- all of which predict that approximately 10\,000 DWDs will be detectable by LISA at high SNR. Using the results from \cite{maoz_separation_2018}, \cite{korol_observationally_2022} estimate that the number of DWD binaries detectable by LISA with SNR $>$ 7 is about 60\,000, and typically about 2-5 times larger than the binary population synthesis calculations. As we discussed earlier, the differences in WD mass distribution can significantly alter the DWD population which might explain the over abundance of LISA detectable DWD binaries they discover. A careful handling of the mass distribution and the volume limited nature of these surveys may alleviate some of the discrepancies.

\section{Acknowledgements}

GAP acknowledges helpful comments from Warren Brown, Na'ama Hallakoun and Dan Maoz. GAP acknowledges partial support by the JHU President's Frontier Award to NLZ. 

Funding for the Sloan Digital Sky Survey V has been provided by the Alfred P. Sloan Foundation, the Heising-Simons Foundation, the National Science Foundation, and the Participating Institutions. SDSS acknowledges support and resources from the Center for High-Performance Computing at the University of Utah. SDSS telescopes are located at Apache Point Observatory, funded by the Astrophysical Research Consortium and operated by New Mexico State University, and at Las Campanas Observatory, operated by the Carnegie Institution for Science. The SDSS web site is \url{www.sdss.org}.

SDSS is managed by the Astrophysical Research Consortium for the Participating Institutions of the SDSS Collaboration, including the Carnegie Institution for Science, Chilean National Time Allocation Committee (CNTAC) ratified researchers, Caltech, the Gotham Participation Group, Harvard University, Heidelberg University, The Flatiron Institute, The Johns Hopkins University, L'Ecole polytechnique f\'{e}d\'{e}rale de Lausanne (EPFL), Leibniz-Institut f\"{u}r Astrophysik Potsdam (AIP), Max-Planck-Institut f\"{u}r Astronomie (MPIA Heidelberg), Max-Planck-Institut f\"{u}r Extraterrestrische Physik (MPE), Nanjing University, National Astronomical Observatories of China (NAOC), New Mexico State University, The Ohio State University, Pennsylvania State University, Smithsonian Astrophysical Observatory, Space Telescope Science Institute (STScI), the Stellar Astrophysics Participation Group, Universidad Nacional Aut\'{o}noma de M\'{e}xico, University of Arizona, University of Colorado Boulder, University of Illinois at Urbana-Champaign, University of Toronto, University of Utah, University of Virginia, Yale University, and Yunnan University.

This research has made use of the NASA/IPAC Infrared Science Archive, which is funded by the National Aeronautics and Space Administration and operated by the California Institute of Technology.

The PanSTARRS data used in this paper is from
\cite{stsci_pan-starrs1_2022} and \cite{stsci_pan-starrs1_2022-1}. GALEX data used in this paper is from \citep{stsci_galexmcat_2013}. The 2MASS and CatWISE datasets is from \cite{skrutskie_2mass_2003} and \cite{marocco_catwise2020_2020}.

\software{astropy \citep{astropy_collaboration_astropy_2013,astropy_collaboration_astropy_2018,astropy_collaboration_astropy_2022}, numpy \citep{harris_array_2020}, scipy \citep{virtanen_scipy_2020}, matplotlib \citep{hunter_matplotlib_2007}, astroquery \citep{ginsburg_astroquery_2019}, simbad \citep{wenger_simbad_2000}}

\bibliography{references}

\begin{thebibliography}{}
\expandafter\ifx\csname natexlab\endcsname\relax\def\natexlab#1{#1}\fi
\providecommand{\url}[1]{\href{#1}{#1}}
\providecommand{\dodoi}[1]{doi:~\href{http://doi.org/#1}{\nolinkurl{#1}}}
\providecommand{\doeprint}[1]{\href{http://ascl.net/#1}{\nolinkurl{http://ascl.net/#1}}}
\providecommand{\doarXiv}[1]{\href{https://arxiv.org/abs/#1}{\nolinkurl{https://arxiv.org/abs/#1}}}

\bibitem[{{Abdurro'uf} {et~al.}(2022){Abdurro'uf}, Accetta, Aerts,
  Silva~Aguirre, Ahumada, Ajgaonkar, Filiz~Ak, Alam, Allende~Prieto, Almeida,
  Anders, Anderson, Andrews, Anguiano, Aquino-Ortíz, Aragón-Salamanca,
  Argudo-Fernández, Ata, Aubert, Avila-Reese, Badenes, Barbá, Barger,
  Barrera-Ballesteros, Beaton, Beers, Belfiore, Bender, Bernardi, Bershady,
  Beutler, Bidin, Bird, Bizyaev, Blanc, Blanton, Boardman, Bolton, Boquien,
  Borissova, Bovy, Brandt, Brown, Brownstein, Brusa, Buchner, Bundy, Burchett,
  Bureau, Burgasser, Cabang, Campbell, Cappellari, Carlberg, Wanderley,
  Carrera, Cash, Chen, Chen, Cherinka, Chiappini, Choi, Chojnowski, Chung,
  Clerc, Cohen, Comerford, Comparat, da~Costa, Covey, Crane, Cruz-Gonzalez,
  Culhane, Cunha, Dai, Damke, Darling, Davidson, Davies, Dawson, De~Lee,
  Diamond-Stanic, Cano-Díaz, Sánchez, Donor, Duckworth, Dwelly, Eisenstein,
  Elsworth, Emsellem, Eracleous, Escoffier, Fan, Farr, Feng,
  Fernández-Trincado, Feuillet, Filipp, Fillingham, Frinchaboy, Fromenteau,
  Galbany, García, García-Hernández, Ge, Geisler, Gelfand, Géron, Gibson,
  Goddy, Godoy-Rivera, Grabowski, Green, Greener, Grier, Griffith, Guo, Guy,
  Hadjara, Harding, Hasselquist, Hayes, Hearty, Hernández, Hill, Hogg,
  Holtzman, Horta, Hsieh, Hsu, Hsu, Huber, Huertas-Company, Hutchinson, Hwang,
  Ibarra-Medel, Chitham, Ilha, Imig, Jaekle, Jayasinghe, Ji, Johnson, Jones,
  Jönsson, Katkov, Khalatyan, Kinemuchi, Kisku, Knapen, Kneib, Kollmeier,
  Kong, Kounkel, Kreckel, Krishnarao, Lacerna, Lane, Langgin, Lavender, Law,
  Lazarz, Leung, Leung, Lewis, Li, Li, Lian, Liang, Lin, Lin, Lin, Lintott,
  Long, Longa-Peña, López-Cobá, Lu, Lundgren, Luo, Mackereth, de~la Macorra,
  Mahadevan, Majewski, Manchado, Mandeville, Maraston, Margalef-Bentabol,
  Masseron, Masters, Mathur, McDermid, Mckay, Merloni, Merrifield, Meszaros,
  Miglio, Di~Mille, Minniti, Minsley, \&
  Monachesi}]{abdurrouf_seventeenth_2022}
{Abdurro'uf}, Accetta, K., Aerts, C., {et~al.} 2022, The Astrophysical Journal
  Supplement Series, 259, 35, \dodoi{10.3847/1538-4365/ac4414}

\bibitem[{Adamane~Pallathadka {et~al.}(2024)Adamane~Pallathadka, Chandra,
  Zakamska, Hwang, Zenati, Hermes, El-Badry, Gänsicke, Morrison, Crumpler, \&
  Arseneau}]{adamane_pallathadka_discovery_2024}
Adamane~Pallathadka, G., Chandra, V., Zakamska, N.~L., {et~al.} 2024, The
  Astrophysical Journal, 968, 42, \dodoi{10.3847/1538-4357/ad3e86}

\bibitem[{Adamane~Pallathadka {et~al.}(2025)Adamane~Pallathadka, Chandra,
  Gansicke, Zakamska, Koester, Zenati, Crumpler, Arseneau, Hermes, Schreiber,
  Stassun, Schwope, El-Badry, Tovmassian, Cunningham, \&
  Morrison}]{adamane_pallathadka_double_2025}
Adamane~Pallathadka, G., Chandra, V., Gansicke, B.~T., {et~al.} 2025, Double
  {White} {Dwarf} {Binaries} in {SDSS}-{V} {DR19} : {The} discovery of a rare
  {DA}+{DQ} white dwarf binary with 31 hour orbital period,  arXiv,
  \dodoi{10.48550/arXiv.2507.11618}

\bibitem[{Almeida {et~al.}(2023)Almeida, Anderson, Argudo-Fernández, Badenes,
  Barger, Barrera-Ballesteros, Bender, Benitez, Besser, Bird, Bizyaev, Blanton,
  Bochanski, Bovy, Brandt, Brownstein, Buchner, Bulbul, Burchett, Díaz,
  Carlberg, Casey, Chandra, Cherinka, Chiappini, Coker, Comparat, Conroy,
  Contardo, Cortes, Covey, Crane, Cunha, Dabbieri, Davidson, Davis,
  De~Andrade~Queiroz, De~Lee, Méndez~Delgado, Demasi, Di~Mille, Donor, Dow,
  Dwelly, Eracleous, Eriksen, Fan, Farr, Frederick, Fries, Frinchaboy,
  Gänsicke, Ge, González~Ávila, Grabowski, Grier, Guiglion, Gupta, Hall,
  Hawkins, Hayes, Hermes, Hernández-García, Hogg, Holtzman, Ibarra-Medel, Ji,
  Jofre, Johnson, Jones, Kinemuchi, Kluge, Koekemoer, Kollmeier, Kounkel,
  Krishnarao, Krumpe, Lacerna, Lago, Laporte, Liu, Liu, Liu, Lopes,
  Macktoobian, Majewski, Malanushenko, Maoz, Masseron, Masters, Matijevic,
  McBride, Medan, Merloni, Morrison, Myers, Mészáros, Negrete, Nidever,
  Nitschelm, Oravetz, Oravetz, Pan, Peng, Pinsonneault, Pogge, Qiu, Ramirez,
  Rix, Rosso, Runnoe, Salvato, Sanchez, Santana, Saydjari, Sayres, Schlaufman,
  Schneider, Schwope, Serna, Shen, Sobeck, Song, Souto, Spoo, Stassun,
  Steinmetz, Straumit, Stringfellow, Sánchez-Gallego, Taghizadeh-Popp, Tayar,
  Thakar, Tissera, Tkachenko, Toledo, Trakhtenbrot, Fernández-Trincado, Troup,
  Trump, Tuttle, Ulloa, Vazquez-Mata, Alfaro, Villanova, Wachter, Weijmans,
  Wheeler, Wilson, Wojno, Wolf, Xue, Ybarra, Zari, \&
  Zasowski}]{almeida_eighteenth_2023}
Almeida, A., Anderson, S.~F., Argudo-Fernández, M., {et~al.} 2023, The
  Astrophysical Journal Supplement Series, 267, 44,
  \dodoi{10.3847/1538-4365/acda98}

\bibitem[{Arseneau {et~al.}(2024)Arseneau, Chandra, Hwang, Zakamska,
  Pallathadka, Crumpler, Hermes, El-Badry, Rix, Stassun, Gänsicke, Brownstein,
  \& Morrison}]{arseneau_measuring_2024}
Arseneau, S., Chandra, V., Hwang, H.-C., {et~al.} 2024, The Astrophysical
  Journal, 963, 17, \dodoi{10.3847/1538-4357/ad2168}

\bibitem[{{Astropy Collaboration} {et~al.}(2013){Astropy Collaboration},
  Robitaille, Tollerud, Greenfield, Droettboom, Bray, Aldcroft, Davis,
  Ginsburg, Price-Whelan, Kerzendorf, Conley, Crighton, Barbary, Muna,
  Ferguson, Grollier, Parikh, Nair, Unther, Deil, Woillez, Conseil, Kramer,
  Turner, Singer, Fox, Weaver, Zabalza, Edwards, Azalee~Bostroem, Burke, Casey,
  Crawford, Dencheva, Ely, Jenness, Labrie, Lim, Pierfederici, Pontzen, Ptak,
  Refsdal, Servillat, \& Streicher}]{astropy_collaboration_astropy_2013}
{Astropy Collaboration}, Robitaille, T.~P., Tollerud, E.~J., {et~al.} 2013,
  Astronomy and Astrophysics, 558, A33, \dodoi{10.1051/0004-6361/201322068}

\bibitem[{{Astropy Collaboration} {et~al.}(2018){Astropy Collaboration},
  Price-Whelan, Sipőcz, Günther, Lim, Crawford, Conseil, Shupe, Craig,
  Dencheva, Ginsburg, VanderPlas, Bradley, Pérez-Suárez, de~Val-Borro,
  Aldcroft, Cruz, Robitaille, Tollerud, Ardelean, Babej, Bach, Bachetti,
  Bakanov, Bamford, Barentsen, Barmby, Baumbach, Berry, Biscani, Boquien,
  Bostroem, Bouma, Brammer, Bray, Breytenbach, Buddelmeijer, Burke, Calderone,
  Cano~Rodríguez, Cara, Cardoso, Cheedella, Copin, Corrales, Crichton,
  D'Avella, Deil, Depagne, Dietrich, Donath, Droettboom, Earl, Erben, Fabbro,
  Ferreira, Finethy, Fox, Garrison, Gibbons, Goldstein, Gommers, Greco,
  Greenfield, Groener, Grollier, Hagen, Hirst, Homeier, Horton, Hosseinzadeh,
  Hu, Hunkeler, Ivezić, Jain, Jenness, Kanarek, Kendrew, Kern, Kerzendorf,
  Khvalko, King, Kirkby, Kulkarni, Kumar, Lee, Lenz, Littlefair, Ma, Macleod,
  Mastropietro, McCully, Montagnac, Morris, Mueller, Mumford, Muna, Murphy,
  Nelson, Nguyen, Ninan, Nöthe, Ogaz, Oh, Parejko, Parley, Pascual, Patil,
  Patil, Plunkett, Prochaska, Rastogi, Reddy~Janga, Sabater, Sakurikar,
  Seifert, Sherbert, Sherwood-Taylor, Shih, Sick, Silbiger, Singanamalla,
  Singer, Sladen, Sooley, Sornarajah, Streicher, Teuben, Thomas, Tremblay,
  Turner, Terrón, van Kerkwijk, de~la Vega, Watkins, Weaver, Whitmore,
  Woillez, Zabalza, \& {Astropy
  Contributors}}]{astropy_collaboration_astropy_2018}
{Astropy Collaboration}, Price-Whelan, A.~M., Sipőcz, B.~M., {et~al.} 2018,
  The Astronomical Journal, 156, 123, \dodoi{10.3847/1538-3881/aabc4f}

\bibitem[{{Astropy Collaboration} {et~al.}(2022){Astropy Collaboration},
  Price-Whelan, Lim, Earl, Starkman, Bradley, Shupe, Patil, Corrales, Brasseur,
  Nöthe, Donath, Tollerud, Morris, Ginsburg, Vaher, Weaver, Tocknell,
  Jamieson, van Kerkwijk, Robitaille, Merry, Bachetti, Günther, Aldcroft,
  Alvarado-Montes, Archibald, Bódi, Bapat, Barentsen, Bazán, Biswas, Boquien,
  Burke, Cara, Cara, Conroy, Conseil, Craig, Cross, Cruz, D'Eugenio, Dencheva,
  Devillepoix, Dietrich, Eigenbrot, Erben, Ferreira, Foreman-Mackey, Fox,
  Freij, Garg, Geda, Glattly, Gondhalekar, Gordon, Grant, Greenfield, Groener,
  Guest, Gurovich, Handberg, Hart, Hatfield-Dodds, Homeier, Hosseinzadeh,
  Jenness, Jones, Joseph, Kalmbach, Karamehmetoglu, Kałuszyński, Kelley,
  Kern, Kerzendorf, Koch, Kulumani, Lee, Ly, Ma, MacBride, Maljaars, Muna,
  Murphy, Norman, O'Steen, Oman, Pacifici, Pascual, Pascual-Granado, Patil,
  Perren, Pickering, Rastogi, Roulston, Ryan, Rykoff, Sabater, Sakurikar,
  Salgado, Sanghi, Saunders, Savchenko, Schwardt, Seifert-Eckert, Shih, Jain,
  Shukla, Sick, Simpson, Singanamalla, Singer, Singhal, Sinha, Sipőcz,
  Spitler, Stansby, Streicher, Šumak, Swinbank, Taranu, Tewary, Tremblay,
  de~Val-Borro, Van~Kooten, Vasović, Verma, de~Miranda~Cardoso, Williams,
  Wilson, Winkel, Wood-Vasey, Xue, Yoachim, Zhang, Zonca, \& {Astropy Project
  Contributors}}]{astropy_collaboration_astropy_2022}
{Astropy Collaboration}, Price-Whelan, A.~M., Lim, P.~L., {et~al.} 2022, The
  Astrophysical Journal, 935, 167, \dodoi{10.3847/1538-4357/ac7c74}

\bibitem[{Badenes \& Maoz(2012)}]{badenes_merger_2012}
Badenes, C., \& Maoz, D. 2012, The Astrophysical Journal, 749, L11,
  \dodoi{10.1088/2041-8205/749/1/L11}

\bibitem[{Badenes {et~al.}(2009)Badenes, Mullally, Thompson, \&
  Lupton}]{badenes_first_2009}
Badenes, C., Mullally, F., Thompson, S.~E., \& Lupton, R.~H. 2009, The
  Astrophysical Journal, 707, 971, \dodoi{10.1088/0004-637X/707/2/971}

\bibitem[{Bellm {et~al.}(2019)Bellm, Kulkarni, Graham, Dekany, Smith, Riddle,
  Masci, Helou, Prince, Adams, Barbarino, Barlow, Bauer, Beck, Belicki, Biswas,
  Blagorodnova, Bodewits, Bolin, Brinnel, Brooke, Bue, Bulla, Burruss, Cenko,
  Chang, Connolly, Coughlin, Cromer, Cunningham, De, Delacroix, Desai, Duev,
  Eadie, Farnham, Feeney, Feindt, Flynn, Franckowiak, Frederick, Fremling,
  Gal-Yam, Gezari, Giomi, Goldstein, Golkhou, Goobar, Groom, Hacopians, Hale,
  Henning, Ho, Hover, Howell, Hung, Huppenkothen, Imel, Ip, Ivezić, Jackson,
  Jones, Juric, Kasliwal, Kaspi, Kaye, Kelley, Kowalski, Kramer, Kupfer,
  Landry, Laher, Lee, Lin, Lin, Lunnan, Giomi, Mahabal, Mao, Miller, Monkewitz,
  Murphy, Ngeow, Nordin, Nugent, Ofek, Patterson, Penprase, Porter, Rauch,
  Rebbapragada, Reiley, Rigault, Rodriguez, Roestel, Rusholme, Santen, Schulze,
  Shupe, Singer, Soumagnac, Stein, Surace, Sollerman, Szkody, Taddia, Terek,
  Van~Sistine, Van~Velzen, Vestrand, Walters, Ward, Ye, Yu, Yan, \&
  Zolkower}]{bellm_zwicky_2019}
Bellm, E.~C., Kulkarni, S.~R., Graham, M.~J., {et~al.} 2019, Publications of
  the Astronomical Society of the Pacific, 131, 018002,
  \dodoi{10.1088/1538-3873/aaecbe}

\bibitem[{Bowen \& Vaughan(1973)}]{bowen_optical_1973}
Bowen, I.~S., \& Vaughan, A.~H. 1973, Applied Optics, 12, 1430,
  \dodoi{10.1364/AO.12.001430}

\bibitem[{Breedt {et~al.}(2017)Breedt, Steeghs, Marsh, Gentile~Fusillo,
  Tremblay, Green, De~Pasquale, Hermes, Gänsicke, Parsons, Bours, Longa-Peña,
  \& Rebassa-Mansergas}]{breedt_using_2017}
Breedt, E., Steeghs, D., Marsh, T.~R., {et~al.} 2017, Monthly Notices of the
  Royal Astronomical Society, 468, 2910, \dodoi{10.1093/mnras/stx430}

\bibitem[{Brown {et~al.}(2011)Brown, Kilic, Brown, \&
  Kenyon}]{brown_binary_2011}
Brown, J.~M., Kilic, M., Brown, W.~R., \& Kenyon, S.~J. 2011, The Astrophysical
  Journal, 730, 67, \dodoi{10.1088/0004-637X/730/2/67}

\bibitem[{Brown {et~al.}(2013)Brown, Kilic, Allende~Prieto, Gianninas, \&
  Kenyon}]{brown_elm_2013}
Brown, W.~R., Kilic, M., Allende~Prieto, C., Gianninas, A., \& Kenyon, S.~J.
  2013, The Astrophysical Journal, 769, 66, \dodoi{10.1088/0004-637X/769/1/66}

\bibitem[{Brown {et~al.}(2016)Brown, Kilic, Kenyon, \&
  Gianninas}]{brown_most_2016}
Brown, W.~R., Kilic, M., Kenyon, S.~J., \& Gianninas, A. 2016, The
  Astrophysical Journal, 824, 46, \dodoi{10.3847/0004-637X/824/1/46}

\bibitem[{Brown {et~al.}(2010)Brown, Kilic, Prieto, \& Kenyon}]{brown_elm_2010}
Brown, W.~R., Kilic, M., Prieto, C.~A., \& Kenyon, S.~J. 2010, The
  Astrophysical Journal, 723, 1072, \dodoi{10.1088/0004-637X/723/2/1072}

\bibitem[{Brown {et~al.}(2020)Brown, Kilic, Kosakowski, Andrews, Heinke,
  Agüeros, Camilo, Gianninas, Hermes, \& Kenyon}]{brown_elm_2020}
Brown, W.~R., Kilic, M., Kosakowski, A., {et~al.} 2020, The Astrophysical
  Journal, 889, 49, \dodoi{10.3847/1538-4357/ab63cd}

\bibitem[{Burdge {et~al.}(2020)Burdge, Coughlin, Fuller, Kaplan, Kulkarni,
  Marsh, Bellm, Dekany, Duev, Graham, Mahabal, Masci, Laher, Riddle, Soumagnac,
  \& Prince}]{burdge_88_2020}
Burdge, K.~B., Coughlin, M.~W., Fuller, J., {et~al.} 2020, The Astrophysical
  Journal Letters, 905, L7, \dodoi{10.3847/2041-8213/abca91}

\bibitem[{Bédard {et~al.}(2020)Bédard, Bergeron, Brassard, \&
  Fontaine}]{bedard_spectral_2020}
Bédard, A., Bergeron, P., Brassard, P., \& Fontaine, G. 2020, The
  Astrophysical Journal, 901, 93, \dodoi{10.3847/1538-4357/abafbe}

\bibitem[{Chambers {et~al.}(2016)Chambers, Magnier, Metcalfe, Flewelling,
  Huber, Waters, Denneau, Draper, Farrow, Finkbeiner, Holmberg, Koppenhoefer,
  Price, Rest, Saglia, Schlafly, Smartt, Sweeney, Wainscoat, Burgett, Chastel,
  Grav, Heasley, Hodapp, Jedicke, Kaiser, Kudritzki, Luppino, Lupton, Monet,
  Morgan, Onaka, Shiao, Stubbs, Tonry, White, Bañados, Bell, Bender, Bernard,
  Boegner, Boffi, Botticella, Calamida, Casertano, Chen, Chen, Cole, Deacon,
  Frenk, Fitzsimmons, Gezari, Gibbs, Goessl, Goggia, Gourgue, Goldman, Grant,
  Grebel, Hambly, Hasinger, Heavens, Heckman, Henderson, Henning, Holman, Hopp,
  Ip, Isani, Jackson, Keyes, Koekemoer, Kotak, Le, Liska, Long, Lucey, Liu,
  Martin, Masci, McLean, Mindel, Misra, Morganson, Murphy, Obaika, Narayan,
  Nieto-Santisteban, Norberg, Peacock, Pier, Postman, Primak, Rae, Rai, Riess,
  Riffeser, Rix, Röser, Russel, Rutz, Schilbach, Schultz, Scolnic, Strolger,
  Szalay, Seitz, Small, Smith, Soderblom, Taylor, Thomson, Taylor, Thakar,
  Thiel, Thilker, Unger, Urata, Valenti, Wagner, Walder, Walter, Watters,
  Werner, Wood-Vasey, \& Wyse}]{chambers_pan-starrs1_2016}
Chambers, K.~C., Magnier, E.~A., Metcalfe, N., {et~al.} 2016, The
  {Pan}-{STARRS1} {Surveys},  arXiv, \dodoi{10.48550/arXiv.1612.05560}

\bibitem[{Chandra {et~al.}(2020)Chandra, Hwang, Zakamska, \&
  Budavári}]{chandra_computational_2020}
Chandra, V., Hwang, H.-C., Zakamska, N.~L., \& Budavári, T. 2020, Monthly
  Notices of the Royal Astronomical Society, 497, 2688,
  \dodoi{10.1093/mnras/staa2165}

\bibitem[{Chandra {et~al.}(2021)Chandra, Hwang, Zakamska, Gaensicke, Hermes,
  Schwope, Badenes, Tovmassian, Bauer, Maoz, Schreiber, Toloza, Inight, Rix, \&
  Brown}]{chandra_99-minute_2021}
Chandra, V., Hwang, H.-C., Zakamska, N.~L., {et~al.} 2021, The Astrophysical
  Journal, 921, 160, \dodoi{10.3847/1538-4357/ac2145}

\bibitem[{Chandrasekhar(1931)}]{chandrasekhar_maximum_1931}
Chandrasekhar, S. 1931, The Astrophysical Journal, 74, 81,
  \dodoi{10.1086/143324}

\bibitem[{Chickles {et~al.}(2025)Chickles, Burdge, Chakraborty, Dhillon,
  Draghis, Munday, Rappaport, Tonry, Bauer, Brown, Castro, Chakrabarty, Dyer,
  El-Badry, Frebel, Furesz, Garbutt, Green, Householder, Hughes, Jarvis, Kara,
  Kennedy, Kerry, Littlefair, McCormac, Mo, Ng, Parsons, Pelisoli, Pike,
  Prince, Ricker, van Roestel, Sahman, Shen, Simcoe, Tremblay, Vanderburg, \&
  Wong}]{chickles_gravitational-wave-detectable_2025}
Chickles, E.~T., Burdge, K.~B., Chakraborty, J., {et~al.} 2025, The
  Astrophysical Journal, 987, 206, \dodoi{10.3847/1538-4357/add34c}

\bibitem[{Crumpler {et~al.}(2024)Crumpler, Chandra, Zakamska,
  Adamane~Pallathadka, Arseneau, Gentile~Fusillo, Hermes, Badenes, Chakraborty,
  Gänsicke, \& Schmidt}]{crumpler_detection_2024}
Crumpler, N.~R., Chandra, V., Zakamska, N.~L., {et~al.} 2024, The Astrophysical
  Journal, 977, 237, \dodoi{10.3847/1538-4357/ad8ddc}

\bibitem[{Crumpler {et~al.}(2025)Crumpler, Chandra, Zakamska,
  Adamane~Pallathadka, Arseneau, Gentile~Fusillo, Hermes, Badenes, Chakraborty,
  Gänsicke, Morrison, Rix, Schmidt, Schwope, \& Stassun}]{crumpler_large_2025}
---. 2025, The Astrophysical Journal, 989, 24, \dodoi{10.3847/1538-4357/ade9a9}

\bibitem[{Culpan {et~al.}(2022)Culpan, Geier, Reindl, Pelisoli,
  Gentile~Fusillo, \& Vorontseva}]{culpan_population_2022}
Culpan, R., Geier, S., Reindl, N., {et~al.} 2022, Astronomy \& Astrophysics,
  662, A40, \dodoi{10.1051/0004-6361/202243337}

\bibitem[{Dobbie {et~al.}(2012)Dobbie, Baxter, Külebi, Parker, Koester,
  Jordan, Lodieu, \& Euchner}]{dobbie_two_2012}
Dobbie, P.~D., Baxter, R., Külebi, B., {et~al.} 2012, Monthly Notices of the
  Royal Astronomical Society, 421, 202,
  \dodoi{10.1111/j.1365-2966.2012.20291.x}

\bibitem[{Geier {et~al.}(2019)Geier, Raddi, Gentile~Fusillo, \&
  Marsh}]{geier_population_2019}
Geier, S., Raddi, R., Gentile~Fusillo, N.~P., \& Marsh, T.~R. 2019, Astronomy
  \& Astrophysics, 621, A38, \dodoi{10.1051/0004-6361/201834236}

\bibitem[{Geier {et~al.}(2017)Geier, Østensen, Nemeth, Gentile~Fusillo,
  Gänsicke, Telting, Green, \& Schaffenroth}]{geier_population_2017}
Geier, S., Østensen, R.~H., Nemeth, P., {et~al.} 2017, Astronomy and
  Astrophysics, 600, A50, \dodoi{10.1051/0004-6361/201630135}

\bibitem[{Gentile~Fusillo {et~al.}(2021)Gentile~Fusillo, Tremblay, Cukanovaite,
  Vorontseva, Lallement, Hollands, Gänsicke, Burdge, McCleery, \&
  Jordan}]{gentile_fusillo_catalogue_2021}
Gentile~Fusillo, N.~P., Tremblay, P.~E., Cukanovaite, E., {et~al.} 2021,
  Monthly Notices of the Royal Astronomical Society, 508, 3877,
  \dodoi{10.1093/mnras/stab2672}

\bibitem[{Gentile Fusillo {et~al.}(2019)Gentile Fusillo, Tremblay, Gänsicke,
  Manser, Cunningham, Cukanovaite, Hollands, Marsh, Raddi, Jordan, Toonen,
  Geier, Barstow, \& Cummings}]{gentilefusillo_gaia_2019}
Gentile Fusillo, N.~P., Tremblay, P.-E., Gänsicke, B.~T., {et~al.} 2019,
  Monthly Notices of the Royal Astronomical Society, 482, 4570,
  \dodoi{10.1093/mnras/sty3016}

\bibitem[{Giammichele {et~al.}(2012)Giammichele, Bergeron, \&
  Dufour}]{giammichele_know_2012}
Giammichele, N., Bergeron, P., \& Dufour, P. 2012, The Astrophysical Journal
  Supplement Series, 199, 29, \dodoi{10.1088/0067-0049/199/2/29}

\bibitem[{Ginsburg {et~al.}(2019)Ginsburg, Sipőcz, Brasseur, Cowperthwaite,
  Craig, Deil, Guillochon, Guzman, Liedtke, Lian~Lim, Lockhart, Mommert,
  Morris, Norman, Parikh, Persson, Robitaille, Segovia, Singer, Tollerud,
  de~Val-Borro, Valtchanov, Woillez, {Astroquery Collaboration}, \& {a subset
  of astropy Collaboration}}]{ginsburg_astroquery_2019}
Ginsburg, A., Sipőcz, B.~M., Brasseur, C.~E., {et~al.} 2019, The Astronomical
  Journal, 157, 98, \dodoi{10.3847/1538-3881/aafc33}

\bibitem[{Gunn {et~al.}(2006)Gunn, Siegmund, Mannery, Owen, Hull, Leger, Carey,
  Knapp, York, Boroski, Kent, Lupton, Rockosi, Evans, Waddell, Anderson, Annis,
  Barentine, Bartoszek, Bastian, Bracker, Brewington, Briegel, Brinkmann,
  Brown, Carr, Czarapata, Drennan, Dombeck, Federwitz, Gillespie, Gonzales,
  Hansen, Harvanek, Hayes, Jordan, Kinney, Klaene, Kleinman, Kron, Kresinski,
  Lee, Limmongkol, Lindenmeyer, Long, Loomis, McGehee, Mantsch, Neilsen,
  Neswold, Newman, Nitta, Peoples, Pier, Prieto, Prosapio, Rivetta, Schneider,
  Snedden, \& Wang}]{gunn_25_2006}
Gunn, J.~E., Siegmund, W.~A., Mannery, E.~J., {et~al.} 2006, The Astronomical
  Journal, 131, 2332, \dodoi{10.1086/500975}

\bibitem[{Gänsicke {et~al.}(2007)Gänsicke, Marsh, \&
  Southworth}]{gansicke_sdssj104341530855582_2007}
Gänsicke, B.~T., Marsh, T.~R., \& Southworth, J. 2007, Monthly Notices of the
  Royal Astronomical Society, 380, L35,
  \dodoi{10.1111/j.1745-3933.2007.00343.x}

\bibitem[{Harris {et~al.}(2020)Harris, Millman, Van Der~Walt, Gommers,
  Virtanen, Cournapeau, Wieser, Taylor, Berg, Smith, Kern, Picus, Hoyer,
  Van~Kerkwijk, Brett, Haldane, Del~Río, Wiebe, Peterson, Gérard-Marchant,
  Sheppard, Reddy, Weckesser, Abbasi, Gohlke, \& Oliphant}]{harris_array_2020}
Harris, C.~R., Millman, K.~J., Van Der~Walt, S.~J., {et~al.} 2020, Nature, 585,
  357, \dodoi{10.1038/s41586-020-2649-2}

\bibitem[{Heber(2016)}]{heber_hot_2016}
Heber, U. 2016, Publications of the Astronomical Society of the Pacific, 128,
  082001, \dodoi{10.1088/1538-3873/128/966/082001}

\bibitem[{Heintz {et~al.}(2022)Heintz, Hermes, El-Badry, Walsh, van Saders,
  Fields, \& Koester}]{heintz_testing_2022}
Heintz, T.~M., Hermes, J.~J., El-Badry, K., {et~al.} 2022, The Astrophysical
  Journal, 934, 148, \dodoi{10.3847/1538-4357/ac78d9}

\bibitem[{Hunter(2007)}]{hunter_matplotlib_2007}
Hunter, J.~D. 2007, Computing in Science \& Engineering, 9, 90,
  \dodoi{10.1109/MCSE.2007.55}

\bibitem[{Inight {et~al.}(2021)Inight, Gänsicke, Breedt, Marsh, Pala, \&
  Raddi}]{inight_towards_2021}
Inight, K., Gänsicke, B.~T., Breedt, E., {et~al.} 2021, Monthly Notices of the
  Royal Astronomical Society, 504, 2420, \dodoi{10.1093/mnras/stab753}

\bibitem[{Ivanova {et~al.}(2020)Ivanova, Justham, \&
  Ricker}]{ivanova_common_2020}
Ivanova, N., Justham, S., \& Ricker, P. 2020, Common envelope evolution,
  version: 20201201 edn., {AAS}-{IOP} astronomy. [2021 collection] (Bristol,
  UK: IOP Publishing), \dodoi{10.1088/2514-3433/abb6f0}

\bibitem[{Kepler {et~al.}(2013)Kepler, Pelisoli, Jordan, Kleinman, Koester,
  Külebi, Peçanha, Castanheira, Nitta, Costa, Winget, Kanaan, \&
  Fraga}]{kepler_magnetic_2013}
Kepler, S.~O., Pelisoli, I., Jordan, S., {et~al.} 2013, Monthly Notices of the
  Royal Astronomical Society, 429, 2934, \dodoi{10.1093/mnras/sts522}

\bibitem[{Kepler {et~al.}(2019)Kepler, Pelisoli, Koester, Reindl, Geier,
  Romero, Ourique, Oliveira, \& Amaral}]{kepler_white_2019}
Kepler, S.~O., Pelisoli, I., Koester, D., {et~al.} 2019, Monthly Notices of the
  Royal Astronomical Society, 486, 2169, \dodoi{10.1093/mnras/stz960}

\bibitem[{Kilic {et~al.}(2020)Kilic, Bergeron, Kosakowski, Brown, Agüeros, \&
  Blouin}]{kilic_100_2020}
Kilic, M., Bergeron, P., Kosakowski, A., {et~al.} 2020, The Astrophysical
  Journal, 898, 84, \dodoi{10.3847/1538-4357/ab9b8d}

\bibitem[{Kilic {et~al.}(2012)Kilic, Brown, Allende~Prieto, Kenyon, Heinke,
  Agüeros, \& Kleinman}]{kilic_elm_2012}
Kilic, M., Brown, W.~R., Allende~Prieto, C., {et~al.} 2012, The Astrophysical
  Journal, 751, 141, \dodoi{10.1088/0004-637X/751/2/141}

\bibitem[{Kilic {et~al.}(2016)Kilic, Brown, Heinke, Gianninas, Benni, \&
  Agüeros}]{kilic_massive_2016}
Kilic, M., Brown, W.~R., Heinke, C.~O., {et~al.} 2016, Monthly Notices of the
  Royal Astronomical Society, 460, 4176, \dodoi{10.1093/mnras/stw1277}

\bibitem[{Kilic {et~al.}(2021)Kilic, Bédard, \& Bergeron}]{kilic_hidden_2021}
Kilic, M., Bédard, A., \& Bergeron, P. 2021, Monthly Notices of the Royal
  Astronomical Society, 502, 4972, \dodoi{10.1093/mnras/stab439}

\bibitem[{Kilic {et~al.}(2013)Kilic, Hermes, Gianninas, Brown, Heinke,
  Agüeros, Chote, Sullivan, Bell, \& Harrold}]{kilic_found_2013}
Kilic, M., Hermes, J.~J., Gianninas, A., {et~al.} 2013, Monthly Notices of the
  Royal Astronomical Society: Letters, 438, L26, \dodoi{10.1093/mnrasl/slt151}

\bibitem[{Kollmeier {et~al.}(2025)Kollmeier, Rix, Aerts, Aird, Alfaro, Almeida,
  Anderson, Jiménez~Arranz, Arseneau, Assef, Aviram, Aydar, Badenes,
  Bandyopadhyay, Barger, Barkhouser, Bauer, Bender, Besser, Bhattarai, Bilgi,
  Bird, Bizyaev, Blanc, Blanton, Bochanski, Bovy, Brandon, Brandt, Brownstein,
  Buchner, Burchett, Carlberg, Casey, Castaneda-Carlos, Chakraborty, Chanamé,
  Chandra, Cherinka, Chilingarian, Comparat, Cosens, Covey, Crane, Crumpler,
  Cunha, Cunningham, Dai, Darling, Davidson, Davis, De~Lee, Deacon,
  Méndez~Delgado, Demasi, Demianenko, Derwent, D'Onghia, Di~Mille, Dias,
  Donor, Drory, Dwelly, Egorov, Egorova, El-Badry, Engelman, Eracleous, Fan,
  Farr, Fries, Frinchaboy, Froning, Gänsicke, García, Gelfand,
  Gentile~Fusillo, Glover, Grabowski, Grebel, Green, Grier, Gupta, Gray,
  Häberle, Hall, Hammond, Hawkins, Harding, Hegedűs, Herbst, Hermes,
  Rodríguez~Hidalgo, Hilder, Hogg, Holtzman, Horta, Huang, Hwang,
  Ibarra-Medel, Imig, Inight, Jana, Ji, Jofre, Johns, Johnson, Johnson,
  Johnston, Jones, Katkov, Koekemoer, Kounkel, Kreckel, Krishnarao, Krumpe,
  Kumari, Kupfer, Lacerna, Laporte, Lepine, Li, Liu, Loebman, Long,
  Roman-Lopes, Lu, Majewski, Maoz, McKinnon, Medan, Merloni, Minniti, Morrison,
  Myers, Mészáros, Nandra, Nayak, Ness, Nidever, O'Brien, Oeur, Oravetz,
  Oravetz, Otto, Adamane~Pallathadka, Palunas, Pan, Pappalardo, Pandey,
  Negrete~Peñaloza, Pinsonneault, Pogge, Taghizadeh~Popp, Price-Whelan,
  Pulatova, Qiu, Ramirez, Rankine, Ricci, Runnoe, Sanchez, Salvato, Sattler,
  Saydjari, Sayres, Schlaufman, Schneider, Schreiber, Schwope, Serna, Shen,
  Sifón, Singh, Sinha, Smee, Song, Souto, Stassun, Steinmetz, Stone-Martinez,
  Stringfellow, Stutz, {José}, {Sá}, {nchez-Gallego}, Tan, Tayar, Thai,
  Thakar, Ting, Tkachenko, Tovmasian, Trakhtenbrot, Fernández-Trincado, Troup,
  Trump, Tuttle, van~der Marel, \& Villanova}]{kollmeier_sloan_2025}
Kollmeier, J.~A., Rix, H.-W., Aerts, C., {et~al.} 2025, Sloan {Digital} {Sky}
  {Survey}-{V}: {Pioneering} {Panoptic} {Spectroscopy},  arXiv,
  \dodoi{10.48550/arXiv.2507.06989}

\bibitem[{Korol {et~al.}(2022)Korol, Hallakoun, Toonen, \&
  Karnesis}]{korol_observationally_2022}
Korol, V., Hallakoun, N., Toonen, S., \& Karnesis, N. 2022, Monthly Notices of
  the Royal Astronomical Society, 511, 5936, \dodoi{10.1093/mnras/stac415}

\bibitem[{Kosakowski {et~al.}(2025)Kosakowski, Dorsch, Brown, Kupfer, Ben~Daya,
  \& Kilic}]{kosakowski_new_2025}
Kosakowski, A., Dorsch, M., Brown, W.~R., {et~al.} 2025, The Astrophysical
  Journal, 987, 205, \dodoi{10.3847/1538-4357/add1cf}

\bibitem[{Kupfer {et~al.}(2018)Kupfer, Korol, Shah, Nelemans, Marsh, Ramsay,
  Groot, Steeghs, \& Rossi}]{kupfer_lisa_2018}
Kupfer, T., Korol, V., Shah, S., {et~al.} 2018, Monthly Notices of the Royal
  Astronomical Society, 480, 302, \dodoi{10.1093/mnras/sty1545}

\bibitem[{Lamberts {et~al.}(2019)Lamberts, Blunt, Littenberg, Garrison-Kimmel,
  Kupfer, \& Sanderson}]{lamberts_predicting_2019}
Lamberts, A., Blunt, S., Littenberg, T.~B., {et~al.} 2019, Monthly Notices of
  the Royal Astronomical Society, 490, 5888, \dodoi{10.1093/mnras/stz2834}

\bibitem[{Li {et~al.}(2019)Li, Chen, Chen, \& Han}]{li_formation_2019}
Li, Z., Chen, X., Chen, H.-L., \& Han, Z. 2019, The Astrophysical Journal, 871,
  148, \dodoi{10.3847/1538-4357/aaf9a1}

\bibitem[{Li {et~al.}(2020)Li, Chen, Chen, Li, Yu, \&
  Han}]{li_gravitational-wave_2020}
Li, Z., Chen, X., Chen, H.-L., {et~al.} 2020, The Astrophysical Journal, 893,
  2, \dodoi{10.3847/1538-4357/ab7dc2}

\bibitem[{Liu {et~al.}(2023)Liu, Hwang, Zakamska, \&
  Thorstensen}]{liu_css160319_2023}
Liu, Y., Hwang, H.-C., Zakamska, N.~L., \& Thorstensen, J.~R. 2023, Monthly
  Notices of the Royal Astronomical Society, 522, 2719,
  \dodoi{10.1093/mnras/stad1156}

\bibitem[{Lomb(1976)}]{lomb_least-squares_1976}
Lomb, N.~R. 1976, Astrophysics and Space Science, 39, 447,
  \dodoi{10.1007/BF00648343}

\bibitem[{Maoz {et~al.}(2012)Maoz, Badenes, \&
  Bickerton}]{maoz_characterizing_2012}
Maoz, D., Badenes, C., \& Bickerton, S.~J. 2012, The Astrophysical Journal,
  751, 143, \dodoi{10.1088/0004-637X/751/2/143}

\bibitem[{Maoz \& Hallakoun(2017)}]{maoz_binary_2017}
Maoz, D., \& Hallakoun, N. 2017, Monthly Notices of the Royal Astronomical
  Society, 467, 1414, \dodoi{10.1093/mnras/stx102}

\bibitem[{Maoz {et~al.}(2018)Maoz, Hallakoun, \&
  Badenes}]{maoz_separation_2018}
Maoz, D., Hallakoun, N., \& Badenes, C. 2018, Monthly Notices of the Royal
  Astronomical Society, 476, 2584, \dodoi{10.1093/mnras/sty339}

\bibitem[{Maoz \& Mannucci(2012)}]{maoz_type-ia_2012}
Maoz, D., \& Mannucci, F. 2012, Publications of the Astronomical Society of
  Australia, 29, 447, \dodoi{10.1071/AS11052}

\bibitem[{Maoz {et~al.}(2014)Maoz, Mannucci, \&
  Nelemans}]{maoz_observational_2014}
Maoz, D., Mannucci, F., \& Nelemans, G. 2014, Annual Review of Astronomy and
  Astrophysics, 52, 107, \dodoi{10.1146/annurev-astro-082812-141031}

\bibitem[{Marocco {et~al.}(2020)Marocco, Eisenhardt, Fowler, Kirkpatrick,
  Meisner, Schlafly, Stanford, Garcia, Caselden, Cushing, Cutri, Faherty,
  Gelino, Gonzalez, Jarrett, Koontz, Mainzer, Marchese, Mobasher, Schlegel,
  Stern, Teplitz, \& Wright}]{marocco_catwise2020_2020}
Marocco, F., Eisenhardt, P. R.~M., Fowler, J.~W., {et~al.} 2020, {CatWISE2020}
  {Catalog},  IPAC, \dodoi{10.26131/IRSA551}

\bibitem[{Marocco {et~al.}(2021)Marocco, Eisenhardt, Fowler, Kirkpatrick,
  Meisner, Schlafly, Stanford, Garcia, Caselden, Cushing, Cutri, Faherty,
  Gelino, Gonzalez, Jarrett, Koontz, Mainzer, Marchese, Mobasher, Schlegel,
  Stern, Teplitz, \& Wright}]{marocco_catwise2020_2021}
---. 2021, The Astrophysical Journal Supplement Series, 253, 8,
  \dodoi{10.3847/1538-4365/abd805}

\bibitem[{Marsh(2011)}]{marsh_double_2011}
Marsh, T.~R. 2011, Classical and Quantum Gravity, 28, 094019,
  \dodoi{10.1088/0264-9381/28/9/094019}

\bibitem[{Marsh {et~al.}(2004)Marsh, Nelemans, \& Steeghs}]{marsh_mass_2004}
Marsh, T.~R., Nelemans, G., \& Steeghs, D. 2004, Monthly Notices of the Royal
  Astronomical Society, 350, 113, \dodoi{10.1111/j.1365-2966.2004.07564.x}

\bibitem[{Martin {et~al.}(2005)Martin, Fanson, Schiminovich, Morrissey,
  Friedman, Barlow, Conrow, Grange, Jelinsky, Milliard, Siegmund, Bianchi,
  Byun, Donas, Forster, Heckman, Lee, Madore, Malina, Neff, Rich, Small,
  Surber, Szalay, Welsh, \& Wyder}]{martin_galaxy_2005}
Martin, D.~C., Fanson, J., Schiminovich, D., {et~al.} 2005, The Astrophysical
  Journal, 619, L1, \dodoi{10.1086/426387}

\bibitem[{Maxted {et~al.}(2002{\natexlab{a}})Maxted, Marsh, \&
  Moran}]{maxted_mass_2002}
Maxted, P. F.~L., Marsh, T.~R., \& Moran, C. K.~J. 2002{\natexlab{a}}, Monthly
  Notices of the Royal Astronomical Society, 332, 745,
  \dodoi{10.1046/j.1365-8711.2002.05368.x}

\bibitem[{Maxted {et~al.}(2002{\natexlab{b}})Maxted, Marsh, \&
  Moran}]{maxted_radial_2002}
---. 2002{\natexlab{b}}, Monthly Notices of the Royal Astronomical Society,
  319, 305, \dodoi{10.1046/j.1365-8711.2000.03840.x}

\bibitem[{Munday {et~al.}(2024)Munday, Pelisoli, Tremblay, Marsh, Nelemans,
  Bédard, Toonen, Breedt, Cunningham, O’Brien, \& Dawson}]{munday_dbl_2024}
Munday, J., Pelisoli, I., Tremblay, P.~E., {et~al.} 2024, Monthly Notices of
  the Royal Astronomical Society, 532, 2534, \dodoi{10.1093/mnras/stae1645}

\bibitem[{Munday {et~al.}(2025)Munday, Pakmor, Pelisoli, Jones, Sahu, Tremblay,
  Rajamuthukumar, Nelemans, Magee, Toonen, Bédard, \&
  Cunningham}]{munday_super-chandrasekhar_2025}
Munday, J., Pakmor, R., Pelisoli, I., {et~al.} 2025, Nature Astronomy, 9, 872,
  \dodoi{10.1038/s41550-025-02528-4}

\bibitem[{Napiwotzki {et~al.}(2002)Napiwotzki, Koester, Nelemans, Yungelson,
  Christlieb, Renzini, Reimers, Drechsel, \&
  Leibundgut}]{napiwotzki_binaries_2002}
Napiwotzki, R., Koester, D., Nelemans, G., {et~al.} 2002, Astronomy \&
  Astrophysics, 386, 957, \dodoi{10.1051/0004-6361:20020361}

\bibitem[{Napiwotzki {et~al.}(2020)Napiwotzki, Karl, Lisker, Catalán,
  Drechsel, Heber, Homeier, Koester, Leibundgut, Marsh, Moehler, Nelemans,
  Reimers, Renzini, Ströer, \& Yungelson}]{napiwotzki_eso_2020}
Napiwotzki, R., Karl, C.~A., Lisker, T., {et~al.} 2020, Astronomy \&
  Astrophysics, 638, A131, \dodoi{10.1051/0004-6361/201629648}

\bibitem[{Nelemans {et~al.}(2001{\natexlab{a}})Nelemans, Portegies~Zwart,
  Verbunt, \& Yungelson}]{nelemans_population_2001-1}
Nelemans, G., Portegies~Zwart, S.~F., Verbunt, F., \& Yungelson, L.~R.
  2001{\natexlab{a}}, Astronomy and Astrophysics, 368, 939,
  \dodoi{10.1051/0004-6361:20010049}

\bibitem[{Nelemans {et~al.}(2000)Nelemans, Verbunt, Yungelson, \&
  Portegies~Zwart}]{nelemans_reconstructing_2000}
Nelemans, G., Verbunt, F., Yungelson, L.~R., \& Portegies~Zwart, S.~F. 2000,
  Astronomy and Astrophysics, 360, 1011,
  \dodoi{10.48550/arXiv.astro-ph/0006216}

\bibitem[{Nelemans {et~al.}(2001{\natexlab{b}})Nelemans, Yungelson, \&
  Portegies~Zwart}]{nelemans_gravitational_2001}
Nelemans, G., Yungelson, L.~R., \& Portegies~Zwart, S.~F. 2001{\natexlab{b}},
  Astronomy \& Astrophysics, 375, 890, \dodoi{10.1051/0004-6361:20010683}

\bibitem[{Nelemans {et~al.}(2001{\natexlab{c}})Nelemans, Yungelson,
  Portegies~Zwart, \& Verbunt}]{nelemans_population_2001}
Nelemans, G., Yungelson, L.~R., Portegies~Zwart, S.~F., \& Verbunt, F.
  2001{\natexlab{c}}, Astronomy \& Astrophysics, 365, 491,
  \dodoi{10.1051/0004-6361:20000147}

\bibitem[{Parsons {et~al.}(2012)Parsons, Marsh, Gänsicke, Dhillon,
  Copperwheat, Littlefair, Pyrzas, Drake, Koester, Schreiber, \&
  Rebassa-Mansergas}]{parsons_shortest_2012}
Parsons, S.~G., Marsh, T.~R., Gänsicke, B.~T., {et~al.} 2012, Monthly Notices
  of the Royal Astronomical Society, 419, 304,
  \dodoi{10.1111/j.1365-2966.2011.19691.x}

\bibitem[{Pelisoli {et~al.}(2018)Pelisoli, Kepler, \&
  Koester}]{pelisoli_sda_2018}
Pelisoli, I., Kepler, S.~O., \& Koester, D. 2018, Monthly Notices of the Royal
  Astronomical Society, 475, 2480, \dodoi{10.1093/mnras/sty011}

\bibitem[{Perlmutter {et~al.}(1999)Perlmutter, Aldering, Goldhaber, Knop,
  Nugent, Castro, Deustua, Fabbro, Goobar, Groom, Hook, Kim, Kim, Lee, Nunes,
  Pain, Pennypacker, Quimby, Lidman, Ellis, Irwin, McMahon, Ruiz‐Lapuente,
  Walton, Schaefer, Boyle, Filippenko, Matheson, Fruchter, Panagia, Newberg,
  Couch, \& Project}]{perlmutter_measurements_1999}
Perlmutter, S., Aldering, G., Goldhaber, G., {et~al.} 1999, The Astrophysical
  Journal, 517, 565, \dodoi{10.1086/307221}

\bibitem[{Rebassa-Mansergas {et~al.}(2010)Rebassa-Mansergas, Gänsicke,
  Schreiber, Koester, \& Rodríguez-Gil}]{rebassa-mansergas_post-common_2010}
Rebassa-Mansergas, A., Gänsicke, B.~T., Schreiber, M.~R., Koester, D., \&
  Rodríguez-Gil, P. 2010, Monthly Notices of the Royal Astronomical Society,
  402, 620, \dodoi{10.1111/j.1365-2966.2009.15915.x}

\bibitem[{Riess {et~al.}(1998)Riess, Filippenko, Challis, Clocchiatti, Diercks,
  Garnavich, Gilliland, Hogan, Jha, Kirshner, Leibundgut, Phillips, Reiss,
  Schmidt, Schommer, Smith, Spyromilio, Stubbs, Suntzeff, \&
  Tonry}]{riess_observational_1998}
Riess, A.~G., Filippenko, A.~V., Challis, P., {et~al.} 1998, The Astronomical
  Journal, 116, 1009, \dodoi{10.1086/300499}

\bibitem[{Robson {et~al.}(2019)Robson, Cornish, \&
  Liu}]{robson_construction_2019}
Robson, T., Cornish, N.~J., \& Liu, C. 2019, Classical and Quantum Gravity, 36,
  105011, \dodoi{10.1088/1361-6382/ab1101}

\bibitem[{Romero {et~al.}(2022)Romero, Kepler, Hermes, Amaral, Uzundag,
  Bognár, Bell, VanWyngarden, Baran, Pelisoli, Oliveira, Koester, Klippel,
  Fraga, Bradley, Vučković, Heintz, Reding, Kaiser, \&
  Charpinet}]{romero_discovery_2022}
Romero, A.~D., Kepler, S.~O., Hermes, J.~J., {et~al.} 2022, Monthly Notices of
  the Royal Astronomical Society, 511, 1574, \dodoi{10.1093/mnras/stac093}

\bibitem[{Ruiter {et~al.}(2010)Ruiter, Belczynski, Benacquista, Larson, \&
  Williams}]{ruiter_lisa_2010}
Ruiter, A.~J., Belczynski, K., Benacquista, M., Larson, S.~L., \& Williams, G.
  2010, The Astrophysical Journal, 717, 1006,
  \dodoi{10.1088/0004-637X/717/2/1006}

\bibitem[{Scargle(1982)}]{scargle_studies_1982}
Scargle, J.~D. 1982, The Astrophysical Journal, 263, 835,
  \dodoi{10.1086/160554}

\bibitem[{{SDSS Collaboration} {et~al.}(2025){SDSS Collaboration},
  Adamane~Pallathadka, Aghakhanloo, Aird, Almeida, Amrita, Anders, Anderson,
  Arseneau, González~Avila, Aviram, Aydar, Badenes, Barrera-Ballesteros,
  Bauer, Behmard, Berg, Besser, Moni~Bidin, Bizyaev, Blanc, Blanton, Bovy,
  Brandt, Brownstein, Buchner, Bulbul, Burchett, Carigi, Carlberg, Casey,
  Chakraborty, Chanamé, Chandra, Chiappini, Chilingarian, Comparat, Covey,
  Crumpler, Cunha, D'Onghia, Dai, Darling, Davis, De~Lee, Deacon,
  Méndez~Delgado, Demasi, Demianenko, Demke, Donor, Drory, Villa~Durango,
  Dwelly, Egorov, Egorova, El-Badry, Eracleous, Fan, Farr, Finkbeiner, Fries,
  Frinchaboy, Gentile~Fusillo, Serrano~Félix, Gaensicke, Galligan, García,
  Gelfand, Grabowski, Grebel, Green, Greve, Grier, Griffith, Guetzoyan, Gupta,
  Hackshaw, Hall, Hawkins, Hegedűs, Hekker, Herbst, Hermes,
  Hernández-García, Hiremath, Hogg, Holtzman, Horne, Horta, Huang,
  Hutchinson, Häberle, Ibarra-Medel, Ji, Jofre, Johnson, Johnson, Johnston,
  Kaldor, Katkov, Khalatyan, Khoperskov, Klessen, Kluge, Koekemoer, Kollmeier,
  Kounkel, Kreckel, Krishnarao, Krumpe, Lacerna, Laporte, Lepine, Li, Liang,
  Limberg, Liu, Loebman, Long, Lu, Lucey, Lugo-Aranda, Martínez
  Martinez-Aldama, McKinnon, Medan, Merloni, Morrison, Myers, Mészáros,
  Müller-Horn, Nepal, Ness, Nidever, Nitschelm, Oravetz, Otto, Pan,
  Pérez~Paolino, Negrete~Peñaloza, Pinsonneault, Taghizadeh~Popp,
  Price-Whelan, Pulatova, Queiroz, Raddick, Rankine, Rix, Román-Zúñiga,
  Fernández~Rosso, Runnoe, Mahmud~Saad, Salvato, Sanchez, Sattler, Saydjari,
  Sayres, Schlaufman, Schneider, Schwope, Seaton, Seeburger, Serna, Sharma,
  Shen, Sinha, Sizemore, Sniegowska, Song, Souto, Stassun, Steinmetz, Stone,
  Stone-Martinez, Stringfellow, Mata~Sánchez, Sánchez-Gallego, Tan, Tayar,
  Thai, Thakar, Thibodeaux, Ting, Tkachenko, Trakhtenbrot, Fernandez~Trincado,
  Troup, Trump, Ulloa, Van~der Marel, Vera, Villanova, Villaseñor, Wang, Way,
  Weijmans, Wheeler, Wilson, Wofford, \&
  Wong}]{sdss_collaboration_nineteenth_2025}
{SDSS Collaboration}, Adamane~Pallathadka, G., Aghakhanloo, M., {et~al.} 2025,
  The {Nineteenth} {Data} {Release} of the {Sloan} {Digital} {Sky} {Survey},
  arXiv, \dodoi{10.48550/arXiv.2507.07093}

\bibitem[{Shen(2015)}]{shen_every_2015}
Shen, K.~J. 2015, The Astrophysical Journal, 805, L6,
  \dodoi{10.1088/2041-8205/805/1/L6}

\bibitem[{Shen {et~al.}(2024)Shen, Boos, \& Townsley}]{shen_almost_2024}
Shen, K.~J., Boos, S.~J., \& Townsley, D.~M. 2024, The Astrophysical Journal,
  975, 127, \dodoi{10.3847/1538-4357/ad7379}

\bibitem[{Shen {et~al.}(2018)Shen, Kasen, Miles, \&
  Townsley}]{shen_sub-chandrasekhar-mass_2018}
Shen, K.~J., Kasen, D., Miles, B.~J., \& Townsley, D.~M. 2018, The
  Astrophysical Journal, 854, 52, \dodoi{10.3847/1538-4357/aaa8de}

\bibitem[{Silvestri {et~al.}(2006)Silvestri, Hawley, West, Szkody, Bochanski,
  Eisenstein, McGehee, Schmidt, Smith, Wolfe, Harris, Kleinman, Liebert, Nitta,
  Barentine, Brewington, Brinkmann, Harvanek, Krzesiński, Long, Neilsen,
  Schneider, \& Snedden}]{silvestri_catalog_2006}
Silvestri, N.~M., Hawley, S.~L., West, A.~A., {et~al.} 2006, The Astronomical
  Journal, 131, 1674, \dodoi{10.1086/499494}

\bibitem[{Silvestri {et~al.}(2007)Silvestri, Lemagie, Hawley, West, Schmidt,
  Liebert, Szkody, Mannikko, Wolfe, Barentine, Brewington, Harvanek,
  Krzesinski, Long, Schneider, \& Snedden}]{silvestri_new_2007}
Silvestri, N.~M., Lemagie, M.~P., Hawley, S.~L., {et~al.} 2007, The
  Astronomical Journal, 134, 741, \dodoi{10.1086/519242}

\bibitem[{Skrutskie {et~al.}(2003)Skrutskie, Cutri, Stiening, Weinberg,
  Schneider, Carpenter, Beichman, Capps, Chester, Elias, Huchra, Liebert,
  Lonsdale, Monet, Price, Seitzer, Jarrett, Kirkpatrick, Gizis, Howard, Evans,
  Fowler, Fullmer, Hurt, Light, Kopan, Marsh, McCallon, Tam, Van~Dyk, \&
  Wheelock}]{skrutskie_2mass_2003}
Skrutskie, M.~F., Cutri, R.~M., Stiening, R., {et~al.} 2003, {2MASS}
  {All}-{Sky} {Point} {Source} {Catalog},  IPAC, \dodoi{10.26131/IRSA2}

\bibitem[{Skrutskie {et~al.}(2006)Skrutskie, Cutri, Stiening, Weinberg,
  Schneider, Carpenter, Beichman, Capps, Chester, Elias, Huchra, Liebert,
  Lonsdale, Monet, Price, Seitzer, Jarrett, Kirkpatrick, Gizis, Howard, Evans,
  Fowler, Fullmer, Hurt, Light, Kopan, Marsh, McCallon, Tam, Van~Dyk, \&
  Wheelock}]{skrutskie_two_2006}
---. 2006, The Astronomical Journal, 131, 1163, \dodoi{10.1086/498708}

\bibitem[{Smee {et~al.}(2013)Smee, Gunn, Uomoto, Roe, Schlegel, Rockosi, Carr,
  Leger, Dawson, Olmstead, Brinkmann, Owen, Barkhouser, Honscheid, Harding,
  Long, Lupton, Loomis, Anderson, Annis, Bernardi, Bhardwaj, Bizyaev, Bolton,
  Brewington, Briggs, Burles, Burns, Castander, Connolly, Davenport, Ebelke,
  Epps, Feldman, Friedman, Frieman, Heckman, Hull, Knapp, Lawrence, Loveday,
  Mannery, Malanushenko, Malanushenko, Merrelli, Muna, Newman, Nichol, Oravetz,
  Pan, Pope, Ricketts, Shelden, Sandford, Siegmund, Simmons, Smith, Snedden,
  Schneider, SubbaRao, Tremonti, Waddell, \& York}]{smee_multi-object_2013}
Smee, S.~A., Gunn, J.~E., Uomoto, A., {et~al.} 2013, The Astronomical Journal,
  146, 32, \dodoi{10.1088/0004-6256/146/2/32}

\bibitem[{Solheim(2010)}]{solheim_am_2010}
Solheim, J.-E. 2010, Publications of the Astronomical Society of the Pacific,
  122, 1133, \dodoi{10.1086/656680}

\bibitem[{{STScI}(2013)}]{stsci_galexmcat_2013}
{STScI}. 2013, {GALEX}/{MCAT},  STScI/MAST, \dodoi{10.17909/T9H59D}

\bibitem[{{STScI}(2022{\natexlab{a}})}]{stsci_pan-starrs1_2022}
---. 2022{\natexlab{a}}, Pan-{STARRS1} {DR1} {Catalog},  STScI/MAST,
  \dodoi{10.17909/55E7-5X63}

\bibitem[{{STScI}(2022{\natexlab{b}})}]{stsci_pan-starrs1_2022-1}
---. 2022{\natexlab{b}}, Pan-{STARRS1} {DR2} {Catalog},  STScI/MAST,
  \dodoi{10.17909/S0ZG-JX37}

\bibitem[{Toonen {et~al.}(2012)Toonen, Nelemans, \&
  Portegies~Zwart}]{toonen_supernova_2012}
Toonen, S., Nelemans, G., \& Portegies~Zwart, S. 2012, Astronomy \&
  Astrophysics, 546, A70, \dodoi{10.1051/0004-6361/201218966}

\bibitem[{Tovmassian {et~al.}(2010)Tovmassian, Yungelson, Rauch, Suleimanov,
  Napiwotzki, Stasińska, Tomsick, Wilms, Morisset, Peña, \&
  Richer}]{tovmassian_double-degenerate_2010}
Tovmassian, G., Yungelson, L., Rauch, T., {et~al.} 2010, The Astrophysical
  Journal, 714, 178, \dodoi{10.1088/0004-637X/714/1/178}

\bibitem[{Tremblay {et~al.}(2015)Tremblay, Gianninas, Kilic, Ludwig, Steffen,
  Freytag, \& Hermes}]{tremblay_3d_2015}
Tremblay, P.-E., Gianninas, A., Kilic, M., {et~al.} 2015, The Astrophysical
  Journal, 809, 148, \dodoi{10.1088/0004-637X/809/2/148}

\bibitem[{Tremblay {et~al.}(2013)Tremblay, Ludwig, Steffen, \&
  Freytag}]{tremblay_spectroscopic_2013}
Tremblay, P.-E., Ludwig, H.-G., Steffen, M., \& Freytag, B. 2013, Astronomy \&
  Astrophysics, 559, A104, \dodoi{10.1051/0004-6361/201322318}

\bibitem[{Van Der~Sluys {et~al.}(2006)Van Der~Sluys, Verbunt, \&
  Pols}]{van_der_sluys_modelling_2006}
Van Der~Sluys, M.~V., Verbunt, F., \& Pols, O.~R. 2006, Astronomy \&
  Astrophysics, 460, 209, \dodoi{10.1051/0004-6361:20065066}

\bibitem[{Virtanen {et~al.}(2020)Virtanen, Gommers, Oliphant, Haberland, Reddy,
  Cournapeau, Burovski, Peterson, Weckesser, Bright, Van Der~Walt, Brett,
  Wilson, Millman, Mayorov, Nelson, Jones, Kern, Larson, Carey, Polat, Feng,
  Moore, VanderPlas, Laxalde, Perktold, Cimrman, Henriksen, Quintero, Harris,
  Archibald, Ribeiro, Pedregosa, Van~Mulbregt, {SciPy 1.0 Contributors},
  Vijaykumar, Bardelli, Rothberg, Hilboll, Kloeckner, Scopatz, Lee, Rokem,
  Woods, Fulton, Masson, Häggström, Fitzgerald, Nicholson, Hagen, Pasechnik,
  Olivetti, Martin, Wieser, Silva, Lenders, Wilhelm, Young, Price, Ingold,
  Allen, Lee, Audren, Probst, Dietrich, Silterra, Webber, Slavič, Nothman,
  Buchner, Kulick, Schönberger, De~Miranda~Cardoso, Reimer, Harrington,
  Rodríguez, Nunez-Iglesias, Kuczynski, Tritz, Thoma, Newville, Kümmerer,
  Bolingbroke, Tartre, Pak, Smith, Nowaczyk, Shebanov, Pavlyk, Brodtkorb, Lee,
  McGibbon, Feldbauer, Lewis, Tygier, Sievert, Vigna, Peterson, More, Pudlik,
  Oshima, Pingel, Robitaille, Spura, Jones, Cera, Leslie, Zito, Krauss,
  Upadhyay, Halchenko, \& Vázquez-Baeza}]{virtanen_scipy_2020}
Virtanen, P., Gommers, R., Oliphant, T.~E., {et~al.} 2020, Nature Methods, 17,
  261, \dodoi{10.1038/s41592-019-0686-2}

\bibitem[{Wenger {et~al.}(2000)Wenger, Ochsenbein, Egret, Dubois, Bonnarel,
  Borde, Genova, Jasniewicz, Laloë, Lesteven, \& Monier}]{wenger_simbad_2000}
Wenger, M., Ochsenbein, F., Egret, D., {et~al.} 2000, Astronomy and
  Astrophysics Supplement Series, 143, 9, \dodoi{10.1051/aas:2000332}

\bibitem[{Wilson {et~al.}(2019)Wilson, Hearty, Skrutskie, Majewski, Holtzman,
  Eisenstein, Gunn, Blank, Henderson, Smee, Nelson, Nidever, Arns, Barkhouser,
  Barr, Beland, Bershady, Blanton, Brunner, Burton, Carey, Carr, Colque, Crane,
  Damke, Davidson, Dean, Di~Mille, Don, Ebelke, Evans, Fitzgerald, Gillespie,
  Hall, Harding, Harding, Hammond, Hancock, Harrison, Hope, Horne, Karakla,
  Lam, Leger, MacDonald, Maseman, Matsunari, Melton, Mitcheltree, O’Brien,
  O’Connell, Patten, Richardson, Rieke, Rieke, Roman-Lopes, Schiavon, Sobeck,
  Stolberg, Stoll, Tembe, Trujillo, Uomoto, Vernieri, Walker, Weinberg, Young,
  Anthony-Brumfield, Bizyaev, Breslauer, Lee, Downey, Halverson, Huehnerhoff,
  Klaene, Leon, Long, Mahadevan, Malanushenko, Nguyen, Owen, Sánchez-Gallego,
  Sayres, Shane, Shectman, Shetrone, Skinner, Stauffer, \&
  Zhao}]{wilson_apache_2019}
Wilson, J.~C., Hearty, F.~R., Skrutskie, M.~F., {et~al.} 2019, Publications of
  the Astronomical Society of the Pacific, 131, 055001,
  \dodoi{10.1088/1538-3873/ab0075}

\bibitem[{Yan {et~al.}(2024)Yan, Zhao, Shi, Guo, Li, Lei, \&
  Zhao}]{yan_search_2024}
Yan, H., Zhao, J., Shi, W., {et~al.} 2024, Astronomy \& Astrophysics, 684,
  A103, \dodoi{10.1051/0004-6361/202347617}

\end{thebibliography}
\bibliographystyle{aasjournal}

\appendix

\section{CV candidates}
\begin{figure}[htb]
    \centering
    \includegraphics[width=0.7\linewidth]{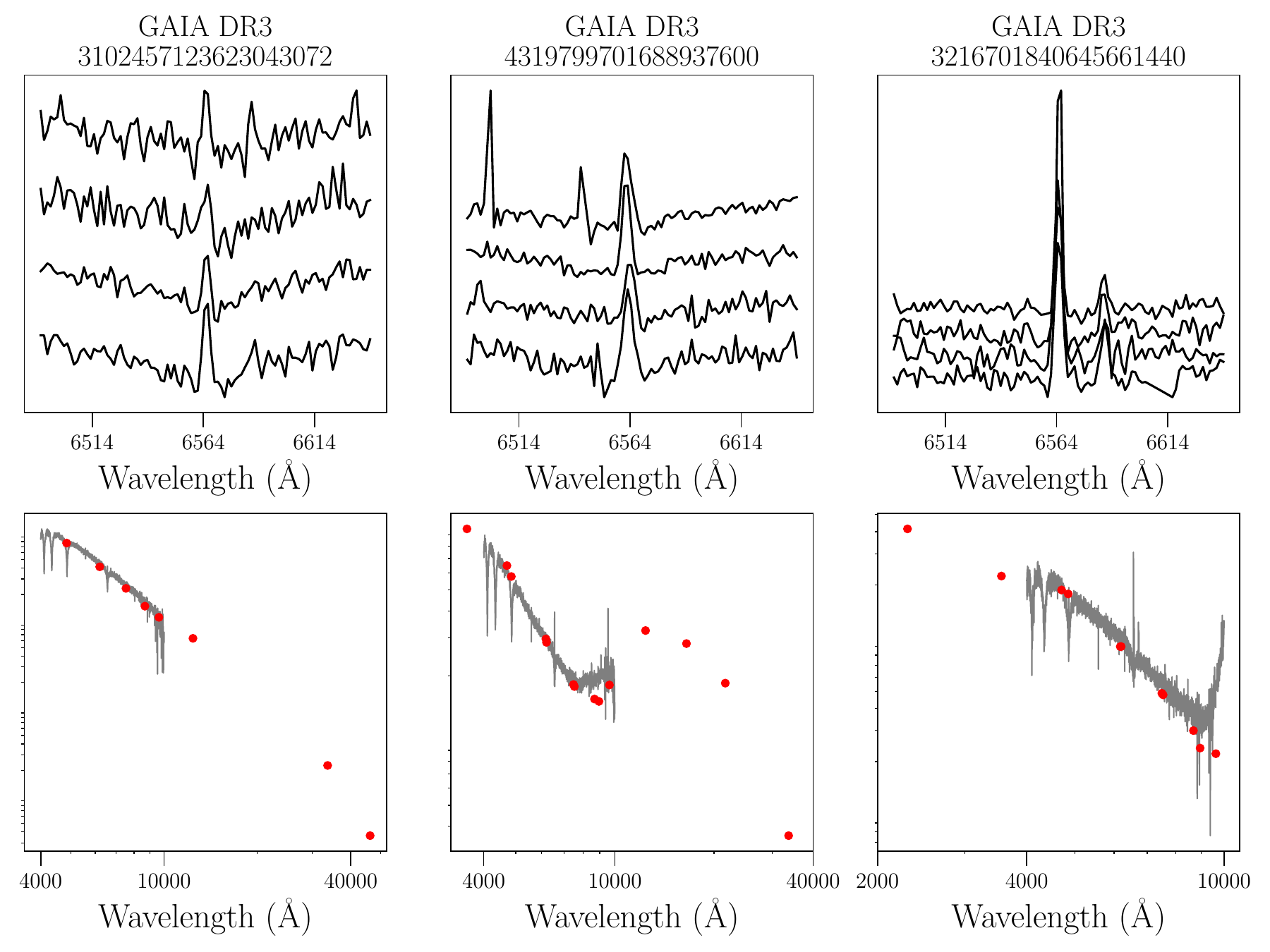}
    \caption{Top: Region around H$\alpha$ line of SDSS spectra are shown for the three mass transferring CV candidates. Bottom: Coadded SDSS spectra and available archival photometry is shown for the same three systems. The absolute flux of the coadded spectra are adjusted to match the photometry. In two of these systems we see a clear infrared excess in either the red end of SDSS spectrum or in infrared photometry, indicating the presence of a mass transferring non-degenerate companion.}
    \label{fig:cv_candidates}
\end{figure}

\begin{figure}[htb]
    \centering
    \includegraphics[width=0.7\linewidth]{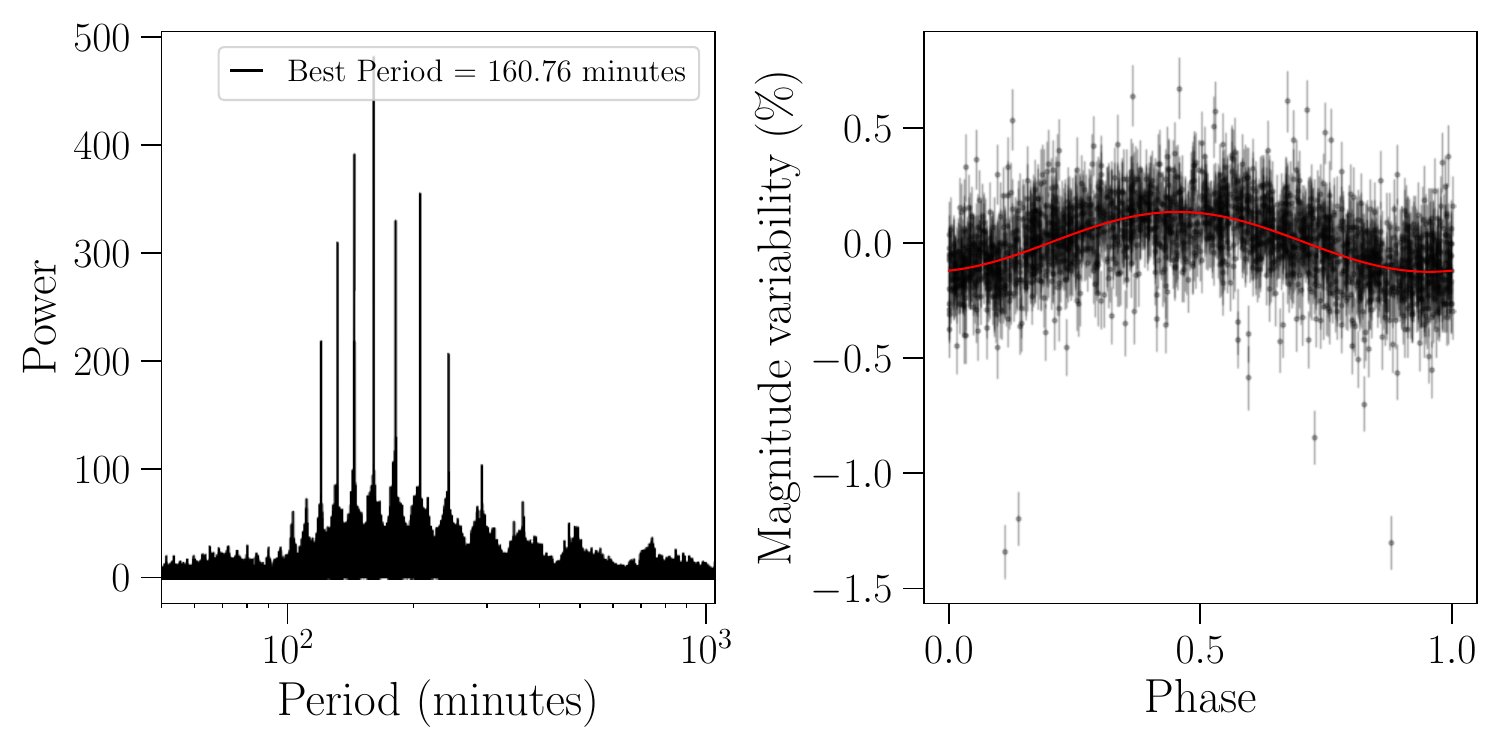}
    \caption{Left: The power spectrum from Lomb-Scargle periodogram of the ZTF lightcurve of Gaia DR3 4319799701688937600. Right: The phase folded ZTF lightcurve is shown along with the best-fit sinusoidal model showing clear periodic variability.}
    \label{fig:CV_cand_4319799701688937600}
\end{figure}

In Fig.~\ref{fig:cv_candidates} we show the SDSS spectrum around H$\alpha$ for the three CV candidates Gaia DR3 3216701840645661440, Gaia DR3 4319799701688937600, and Gaia DR3 3102457123623043072. The presence of emission lines indicates that these systems are accreting matter. Using IRSA, MAST, and \texttt{astroquery}, we collected the available archival photometry for these targets in Galaxy Evolution Explorer \citep[GALEX;][]{martin_galaxy_2005}, SDSS \citep{abdurrouf_seventeenth_2022}, Panoramic Survey Telescope and Rapid Response System \citep[Pan-STARRS;][]{chambers_pan-starrs1_2016}, Two Micron All-Sky Survey \citep[2MASS;][]{skrutskie_two_2006}, and Wide-field Infrared Survey Explorer \citep[WISE;][]{marocco_catwise2020_2021} bands. The coadded SDSS-V along with the available photometry is shown in Fig.~\ref{fig:cv_candidates}. Gaia DR3 3216701840645661440 shows an increase in flux at the red end of the spectrum, likely the presence of a companion which can be confirmed by further observations in infrared bands. Gaia DR3 4319799701688937600 shows an infrared excess in both photometry and in the red end of the SDSS spectrum, which is either an M-Dwarf companion, a dusty disk, or both. We then looked at the latest Zwicky Transient Facility \citep[ZTF;][]{bellm_zwicky_2019} DR23 data for the three systems and performed Lomb-Scargle periodogram to identify any periodic variability. Gaia DR3 4319799701688937600 shows a periodically varying lightcurve with a best-fit period of 160.76 minutes. The power-spectrum and the phase folded lightcurve is shown in Fig.~\ref{fig:CV_cand_4319799701688937600}. The presence of emission lines, infrared excess, and the periodically varying lightcurve hints at this being a polar similar to one discussed by \cite{liu_css160319_2023}. A detailed study with follow-up observations can help uncover the nature of this intriguing system.

\label{appendeix:CV}

\section{Catalog of DWD binary candidates}

\begin{table}
\label{tab:dwd_candidates}
\caption{List of DWD binary candidates from SDSS-V}
\begin{tabular}{ll}
\vspace{3pt}
\hspace*{-2.5cm}
        \begin{tabular}[t]{c|c|c|c}
            \textit{Gaia} & NAME & $\eta$ & \textit{Gaia}\\ 
            DR3 SOURCE ID & & & Mass \\
             & & & (M$_{\odot}$)\\ \hline
3847702561575679232 & J0938+0253 & 92.91 & 0.29 \\
1017136594580182400 & J0852+5142 & 79.06 & 0.36 \\
3076962575704962176 & J0831+0013 & 37.75 & 0.18 \\
3238690771828203520 & J0510+0358 & 34.89 & 0.41 \\
1319676603468508544 & J1557+2823 & 34.57 & 0.35 \\
578539413395848704 & J0857+0342 & 30.06 & 0.19 \\
3620282325265021824 & J1345-0632 & 24.69 & 0.38 \\
649261066445946624 & J0825+1152 & 21.47 & -- \\
3847716202391406976 & J0936+0259 & 15.89 & 0.39 \\
1500004000845782912 & J1337+3952 & 15.36 & 0.25 \\
3669445716389952768 & J1420+0439 & 13.21 & 0.39 \\
1289020673097509760 & J1505+3300 & 12.57 & 0.46 \\
3868927607051816320 & J1039+0818 & 12.54 & 0.31 \\
3213108224329909248 & J0459-0347 & 12.17 & 0.60 \\
1609250784690803584 & J1403+5418 & 11.16 & 0.23 \\
1633145818062780544 & J1741+6526 & 10.91 & -- \\
677695609668436736 & J0817+2351 & 10.78 & 0.29 \\
577257520277310848 & J0906+0223 & 10.70 & 0.40 \\
3213127912460143360 & J0459-0335 & 10.58 & 0.47 \\
3669978532853514368 & J1426+0528 & 10.11 & 0.29 \\
3168905043690276992 & J0736+1622 & 9.46 & 0.26 \\
3213022118825045888 & J0506-0317 & 9.23 & 0.56 \\
2492423226139988864 & J0204-0323 & 8.90 & 0.38 \\
3915026861134449664 & J1123+0956 & 7.88 & 0.86 \\
3656509794586217984 & J1441+0310 & 7.85 & 0.28 \\
784186708135977088 & J1123+4450 & 7.85 & 0.38 \\
2485615050141095424 & J0134-0109 & 7.71 & 0.35 \\
3656616923955045376 & J1441+0351 & 7.67 & -- \\
2494155094392583552 & J0206-0249 & 7.08 & 0.28 \\
3126919298834022528 & J0637+0203 & 6.80 & 0.27 \\
875888795390849664 & J0751+2912 & 6.56 & 0.31 \\
3252609416507834496 & J0412-0202 & 6.24 & 0.90 \\
        \end{tabular}
        \hspace*{-1.5cm}
        \begin{tabular}[t]{c|c|c|c}
            \textit{Gaia} & NAME & $\eta$ & \textit{Gaia}\\ 
            DR3 SOURCE ID & & & Mass \\
             & & & (M$_{\odot}$)\\ \hline
585487364810732160 & J0921+0502 & 5.68 & 0.73 \\
3214584494782707456 & J0509-0226 & 5.61 & 0.26 \\
2502044025898006912 & J0235+0055 & 5.61 & 0.36 \\
4311400051373576704 & J1850+0936 & 5.51 & 0.39 \\
4323058787298127360 & J1931+1756 & 5.27 & 0.65 \\
3148768892681659904 & J0751+0934 & 5.14 & 0.84 \\
3225464231060779904 & J0456-0228 & 4.75 & 0.63 \\
3796709342582100096 & J1128-0116 & 4.71 & 0.33 \\
2542602840887750016 & J0032-0118 & 4.59 & 0.79 \\
3213275251314965632 & J0502-0254 & 4.53 & 0.75 \\
2528745524044330368 & J0034-0245 & 4.41 & 0.41 \\
922729884814569088 & J0803+4235 & 4.09 & 0.31 \\
2643225545153702400 & J2351+0108 & 3.88 & -- \\
579482828732706688 & J0920+0454 & 3.84 & 0.51 \\
1609300262714180352 & J1404+5427 & 3.75 & 0.59 \\
1666750569898974208 & J1415+6230 & 3.47 & 0.33 \\
602421256924272384 & J0833+1135 & 3.46 & 0.35 \\
2530257318172454784 & J0035-0226 & 3.44 & 0.53 \\
3095806242204017152 & J0806+0532 & 3.43 & 0.25 \\
671200102992315264 & J0743+1754 & 3.43 & 0.36 \\
3168900744426560384 & J0736+1618 & 3.41 & 0.95 \\
4013354399299869056 & J1215+2917 & 3.27 & 0.69 \\
3213132825902773376 & J0458-0330 & 3.24 & 0.47 \\
2503044645903634688 & J0234+0212 & 3.23 & 0.59 \\
2542432249083805056 & J0037-0133 & 3.19 & 0.67 \\
1709809163229725312 & J1721+8054 & 3.08 & 0.41 \\
1861112790839753984 & J2020+2841 & 3.08 & 0.68 \\
2260805780286092032 & J1801+7218 & 3.06 & 0.38 \\
879036662822920448 & J0748+3025 & 3.04 & 1.09 \\
2533658451234555136 & J0114-0101 & 3.03 & 0.23 \\
3233555915086790400 & J0452+0423 & 3.01 & 0.60 \\
        \end{tabular}\tabularnewline
\end{tabular}
\end{table}
\end{document}